\documentclass[12pt]{article}
\usepackage{setspace} 
\usepackage{amsmath}
\usepackage{graphicx}
\usepackage{enumerate}
\usepackage{natbib}
\usepackage{geometry}                  
\usepackage{array}      
\usepackage{amssymb}
\usepackage{multirow}
\usepackage{float}
\usepackage{diagbox}
\usepackage{appendix}

\newcolumntype{"}{@{\hskip\tabcolsep\vrule width 1pt\hskip\tabcolsep}}
\usepackage{booktabs}

\usepackage{soul, xcolor} 
\usepackage{amsfonts}
\usepackage{amsthm}
\usepackage{bm}
\usepackage[labelformat=simple]{subcaption}
\usepackage{hyperref}

\newcommand{\norm}[1]{\lVert#1\rVert}
\newcommand{\abs}[1]{\lvert#1\rvert}

\DeclareMathOperator*{\argmin}{arg\,min}

\DeclareMathOperator*{\SURE}{SURE}

\DeclareMathOperator*{\tr}{tr}

\newtheorem{property}{Property}
\newtheorem{theorem}{Theorem}
\newtheorem{lemma}{Lemma}

\newtheorem{definition}{Definition}

\title{Sparse additive models in high dimensions with wavelets}

\author{Sylvain Sardy\\ Department of Mathematics, University of Geneva\\ \texttt{sylvain.sardy@unige.ch}\\ 
Xiaoyu Ma \\ Shandong University and University of Geneva\\
\texttt{maxiaoyu0416@gmail.com} \\
}

\begin{document}

\maketitle

\begin{abstract}

In multivariate regression, when covariates are numerous, it is often reasonable to assume that only a small number of them has predictive information.
In some medical applications for instance, it is believed that only a few genes out of thousands are responsible for cancers. In that case, the aim is not only to propose a good fit, but also to select the relevant covariates (genes).
We propose to perform model selection with additive models in high dimensions (sample size and number of covariates). Our approach is computationally efficient thanks to fast wavelet transforms, it does not rely on cross validation,  and it solves a convex optimization problem for a prescribed penalty parameter, called the quantile universal threshold. We also propose a second rule based on Stein unbiased risk estimation geared towards prediction.
We use  Monte Carlo simulations and real data to compare various methods based on  false discovery rate (FDR), true positive rate (TPR) and mean squared error. Our approach is the only one to handle high dimensions, and has the best FDR--TPR trade-off.

   \end{abstract} 
   
   Keywords:    LASSO, model selection, quantile universal threshold.
 

\section{Introduction}

Multivariate regression  aims at predicting a scalar output $y$ from an input vector of covariates ${\bf x}\in {\mathbb R}^p$.
After collecting measurements about a scientific phenomenon of interest, a data set ${\cal D}^{(n)}=({\bf y}^{(n)}, X^{(n)})$ called \emph{traning set} is built, where 
$n$ is the number of samples,
$X^{(n)} \in {\mathbb R}^{n \times p}$ is the matrix of all collected row-wise input vectors ${\bf x}_i=(x_{i,1}, \ldots, x_{i,p})$ of length $p$, for $i=1,\ldots,n$, and
${\bf y}^{(n)}\in {\mathbb R}^n$ are all corresponding outputs.

The common approach in statistical machine learning assumes a multivariate function $\mu$  maps the input  to its output, and that data are realizations from
a random pair of variables $( {\bf X}, Y)$, where the random vector $ {\bf X}$ comes from a $p$-dimensional law and
\begin{equation}\label{eq:YX}
Y\mid {\bf x}=\mu({\bf x})+\epsilon, \quad {\rm where} \quad \epsilon\sim{\rm N}(0,\sigma^2).
\end{equation}
 Using the information in the training set ${\cal D}^{(n)}$, the goal is to find an estimate $\hat \mu$. To measure the predictive quality of $\hat \mu$, one relies on  an independent data set ${\cal D}^{(n')}=({\bf y}^{(n')}, X^{(n')})$ called \emph{test set}, and calculates the predictive error of $\hat \mu$ by ${\rm PE}(\hat \mu, {\cal D}^{(n')})=\sum_{i=1}^{n'}(y_i-\hat \mu({\bf x}_i))^2$.

Predicting well on the training set does not necessarily translate to predicting well on the test set, however. To illustrate this so-called \emph{over-fitting} phenomenon, consider the case where $p>n$, the rank of $X^{(n)}$ is $n$, and the association $\mu$ is linear, that is
\begin{equation}\label{eq:linmod}
\mu({\bf x})=\mu^{{\rm lin}}_{c,{\boldsymbol \beta}}({\bf x}):=c+\sum_{j=1}^{p} \beta_j x_j,
\end{equation}
 where $c$ and ${\boldsymbol \beta}$ are the $p+1$ parameters.
In that case, by solving the least squares problem
$$
\min_{c, {\boldsymbol \beta}} {\rm PE}(\mu^{\rm lin}_{c,{\boldsymbol \beta}}, {\cal D}^{(n)})=\min_{c, {\boldsymbol \beta}} \| {\bf y}^{(n)} - c {\bf 1} - X^{(n)} {\boldsymbol \beta}\|_2^2,
$$
one can find parameters $\hat c$ and $\hat {\boldsymbol \beta}$ such that, $\mu^{{\rm lin}}_{\hat c,\hat {\boldsymbol \beta}}({\bf x}_i)=y_i$ for all measurements $i=1, \ldots, n$ in the training set. But although the corresponding error is null for the training set (${\rm PE}(\mu^{{\rm lin}}_{\hat c,\hat {\boldsymbol \beta}}, {\cal D}^{(n)})=0$), the predictive error ${\rm PE}(\mu^{{\rm lin}}_{\hat c,\hat {\boldsymbol \beta}}, {\cal D}^{(n')})$ for the test set is not null, and may in fact be very large, in particular if $n$ is small and the noise standard deviation $\sigma$ is large.
To allow some bias with respect to  the training set at the estimation stage, a common approach consists in adding a constraint on $\mu$ through its parameters ${\boldsymbol \beta}$. The amount of constraint is indexed by a parameter often named $\lambda$ and called a \emph{regularization} parameter.
In some applications for instance, practitioners believe  that only a small, but unknown, subset of the $p$-long input vector  ${\bf x}$ contains predictive information on the output. In that situation, it makes sense to constrain $\hat {\boldsymbol \beta}$ to have a few non-zero entries.
Calling
\begin{equation} \label{eq:callS}
{\cal S}=\{j \in \{1,\ldots,p \}:\   x_j \mbox{ has predictive information}\}, 
\end{equation}
  a secondary goal of regression becomes to identify ${\cal S}$,
   the set of inputs with predictive information. For linear models \eqref{eq:linmod}, ${\cal S}$ is equivalently defined as ${\cal S}=\{j \in \{1,\ldots,p \}:\   \beta_j \neq 0\}$.
  Assuming a small cardinality $s=|{\cal S}|$   is often reasonable, for instance in some medical applications, where inputs are thousands of genes among which only a few  are believed to have some effect on the output.
To estimate~${\boldsymbol \beta}$ and consequently ${\cal S}$, owing to its $\ell_1$-sparsity inducing penalty on ${\boldsymbol \beta}$, LASSO   \citep{Tibs:regr:1996} solves
$$
(\hat c, \hat {\boldsymbol \beta})^{\rm LASSO}=\argmin_{(c,{\boldsymbol \beta})} {\rm PE}(\mu^{\rm lin}_{c,{\boldsymbol \beta}}, {\cal D}^{(n)})+\lambda \| {\boldsymbol \beta} \|_1.
$$
LASSO identifies potentially important inputs with   $\hat {\cal S}_{\rm lin}^{\rm LASSO}=\{ j\in \{1,\ldots,p \}: \hat {\boldsymbol \beta}^{\rm LASSO}_j \neq 0 \}$.
Remarquably,  even when $p>n$,  one can have $\hat {\cal S}_{\rm lin}^{\rm LASSO}= {\cal S}$ in certain linear regimes depending on $X^{(n)}$ and the signal-to-noise ratio \citep{BuhlGeer11}, for an appropriate choice of $\lambda$ \citep{CaroNickJairoMe2017}.
In other regimes, one cannot retrieve ${\cal S}$ exactly, but one can aim at low false discovery rate (FDR) along with high true positive rate (TPR), defined by
\begin{equation} \label{eq:FDRTPR}
 {\rm FDR}={\mathbb E}\left (  \frac{| \bar {\cal S} \bigcap \hat {\cal S} |}{|\hat {\cal S} | \lor 1} \right ) \quad {\rm and} \quad
 {\rm TPR}={\mathbb E}\left (  \frac{|{\cal S}  \bigcap \hat {\cal S} |}{|{\cal S} |} \right ).
\end{equation}
Controlling the FDR is the goal of the knockoffs \citep{FDRCAndes2015}.

Retrieving ${\cal S}$ when $\mu$ is not linear may be harder.
For instance, if  $x_1$ is useful for prediction, but through $ |x_1|$, the best linear approximation of the absolute value function by a linear association is the constant function, that is, $c+\beta_1 x_1$ with $\beta_1=0$, making the first input impossible to detect with $\hat {\cal S}_{\rm lin}^{\rm LASSO}$.
To help detect such input entries,  additive models assume that  nonlinear associations may occur in all directions by approximating the underlying association with
\begin{equation}\label{eq:YXadd}
\mu^{\rm add}_{c, \mu_1, \ldots, \mu_p}({\bf x})=c+\sum_{j=1}^p \mu_j(x_j),
\end{equation}
where $\mu_j$'s are  univariate functions. Although not dense in multivariate function space, additive models provide more flexibility than linear models.
To fit a wide range of univariate functions $\mu_j$, including linear and absolute value,
the expansion based approach assumes that each univariate function writes as
\begin{equation}\label{eq:expansion}
\mu_j(x)=\sum_{k=1}^{n} \beta_{j,k} \varphi_k(x), 
\end{equation}
where $\{\varphi_k\}_{k=1}^{n}$ are chosen basis functions and $\{\beta_k\}_{k=1}^{n}$ are their corresponding  unknown coefficients that are estimated from the training set. 
Letting
${\boldsymbol \beta}_j=(\beta_{j,1}, \ldots, \beta_{j,n})\in {\mathbb R}^n$ for $j=1,\ldots,p$, expansion-based additive models are a class of functions of the form
\begin{equation}\label{eq:YXadd*}
\mu^{\rm add}_{c, {\boldsymbol \beta}_1, \ldots, {\boldsymbol \beta}_p}({\bf x})=c+\sum_{j=1}^p \sum_{k=1}^{n} \beta_{j,k} \varphi_k(x_j).
\end{equation}
A well-known choice of basis functions $\{\varphi_k\}_{k=1}^{n}$ are splines \citep{Wahb:spli:1990}, as for instance employed by  \citet{hastie1990generalized} and  more recently by \citet{mgcvjasa,mgcv} with {\tt mgcv}.
These models suffer from two drawbacks however: for each of the $p$  directions, they must build and store an $n\times n$ regression matrice of discretized splines for each $\mu_j$, and they must select a regularization parameter $\lambda_j$, for $j=1,\ldots,p$. 
To perform model selection with additive models, \citet{HAM} and \citet{SAM} use sparsity inducing penalties, the former with two hyperparameters and the latter with one hyperparameter. Both still require storing large matrices. 
Consequently methods like {\tt mgcv} and that of \citet{HAM}  are  computationally prohibitive in high dimensions.
Another class of basis functions  $\{\varphi_k\}_{k=1}^{n}$ are wavelets \citep{Daub92}, which have been employed to fit additive models  in low dimension \citep{AMlet04,waveadditive2018,AnestisAM22}.
As we will see, using wavelets has many advantages: no wavelet matrices are stored and a single regularization parameter $\lambda$ indexes the fitting.

The paper is organized as follows. In section~\ref{sct:def}, we describe the model and our new wavelet-based estimator, we show how to solve the corresponding optimization problem in Section~\ref{subsct:BCR}, and we propose two selection rules for the threshold in Sections~\ref{subsct:QUT} and~\ref{subsct:SURE}.
In section~\ref{sct:simulation}, we perform Monte Carlo simulations, then we use three real data sets to compare various methods. Proofs are postponed to the appendix.
The codes are available on \href{https://github.com/StatisticsL/SRAMlet}{https://github.com/StatisticsL/SRAMlet}.



\section{SRAMlet} \label{sct:def}

We consider wavelets  to write each univariate function $\mu_j$ in~\eqref{eq:expansion} as a linear combination of orthonormal basis functions, for $j=1,\ldots,p$. 
As a result, the corresponding regression matrix ${\cal W}^{(n)}$ can be seen as the concatenation of orthonormal matrices. Two key calculations involving this matrix can easily be performed without building and storing this matrix thanks to \citet{MAllat89}'s $O(n)$ ``pyramid'' algorithm: the analysis operation $({\cal W}^{(n)})^{\rm T} {\bf r}$ and the synthesis operation ${\cal W}^{(n)} {\boldsymbol \beta}$ for any ${\bf r}\in {\mathbb R}^{pn}$ and ${\boldsymbol \beta}\in {\mathbb R}^n$. 
To define ${\cal W}^{(n)}$ for a training set ${\cal D}^{(n)}=({\bf y}^{(n)}, X^{(n)})$, let  $P_j$ be the permutation matrices such that $P_j {\bf x}_j$ orders the $j^{\rm th}$ column of $X^{(n)}$ and, using isometric wavelets for unequally spaced samples \citep{SPBGS99,KerkyacharianPicard04}, let $\Phi$ be an orthonormal wavelet matrix with one father wavelet (that is, the constant function) and $(n-1)$ mother wavelets. 
Then we have that each ${\boldsymbol \mu_j}:=(\mu_j(x_{1}^{j}), \ldots, \mu_j(x_{n}^{j}))=P_j^{\rm T} \Phi {\boldsymbol \beta}_j$, for $j=1,\ldots,p$.
So ${\boldsymbol \mu}^{\rm add}_{c, {\boldsymbol \beta}}:=c{\bf 1}+{\boldsymbol \mu_1}+\ldots+{\boldsymbol \mu_p}=c{\bf 1}+{\cal W}^{(n)} {\boldsymbol \beta}$ with 
\begin{equation}\label{eq:orthoblock}
{\cal W}^{(n)}=[P_1^{\rm T} \Phi  \ldots P_p^{\rm T} \Phi] \quad {\rm and} \quad  {\boldsymbol \beta}=({\boldsymbol \beta}_1, \ldots, {\boldsymbol \beta}_p).
\end{equation}
This results in a model with $np$ wavelet coefficients plus a constant $c$, while the input signal only has  length $n$.
Therefore, regularization is needed.
Owing to the \emph{a priori} belief only a few of the $p$ variables have predictive power, and owing to the sparse wavelet representation of univariate functions $\mu_j$ (most mother wavelet coefficients are essentially zero to approximate a function), we regularize the least squares with a sparsity inducing penalty. Inspired  by square-root LASSO \citep{BCW11} and a selection of $\lambda$ that does not require estimation of the noise variance $\sigma^2$ in~\eqref{eq:YX} \citep{CaroNickJairoMe2017}, we define the \emph{square-root additive models with wavelets} (SRAMlet) estimate as 
\begin{equation}\label{eq:sqrtAMlet}
(\hat c, \hat {\boldsymbol \beta})^{\rm SRAMlet} =  \argmin_{c \in {\mathbb R}, {\boldsymbol \beta}\in {\mathbb R}^{pn}}
\sqrt{{\rm PE}(\mu^{\rm add}_{c,{\boldsymbol \beta}}, {\cal D}^{(n)})}
+  \lambda P({\boldsymbol \beta}), 
\end{equation}
for a positive penalty $\lambda$ and a sparsity inducing penalty $P({\boldsymbol \beta})$. 
The corresponding estimation of the indexes ${\cal S}$ of the relevant covariates defined in~\eqref{eq:callS} is given by 
$$
\hat {\cal S}^{\rm SRAMlet}_{\rm add}=\{j \in \{1,\ldots,p \}:\ \| \hat{\boldsymbol  \beta}^{\rm SRAMlet}_j\| \neq 0 \}.
$$
%

 We discuss in the following section the choice of the penalty between $P({\boldsymbol \beta})=\sum_{j=1}^p \| {\boldsymbol \beta}_j\|_2$ of group-LASSO \citep{Yuan:Lin:mode:2006} and  $P({\boldsymbol \beta})=\| {\boldsymbol \beta}\|_1$ of LASSO. 
 Solving the optimization problem~\eqref{eq:sqrtAMlet} and selecting $\lambda$ for~\eqref{eq:sqrtAMlet} is not trivial.
First we propose an efficient algorithm to solve \eqref{eq:sqrtAMlet} in Section~\ref{subsct:BCR}. Second we propose two selection rules for $\lambda$ in Sections~\ref{subsct:QUT} and ~\ref{subsct:SURE}: one geared towards indentification of the indices  ${\cal S}$ of the relevant inputs, and one towards prediction of the output.

\subsection{Optimization} \label{subsct:BCR}

Since needle selection is based on group sparsity on the vectors ${\boldsymbol \beta}_j$, it seems promising to use the group-LASSO penalty. But due to the fact that here ${\cal W}^{(n)}$ is the concatenation of orthonormal matrices, the following theorem proves that choosing the group-LASSO penalty does not work. 
\begin{theorem} \label{th:sqrtgrpLASSO}
 Consider solving \eqref{eq:sqrtAMlet} for a fixed $\lambda \geq 0$,  $P({\boldsymbol \beta})=\|{\boldsymbol \beta}\|_2$ and $\sqrt{{\rm PE}(\mu^{\rm add}_{c,{\boldsymbol \beta}}, {\cal D}^{(n)})}=\|{\bf y}^{(n)}-X^{(n)} {\boldsymbol \beta} \|_2$. Then when the regression matrix $X^{(n)}$ is orthonormal, the solution $\hat {\boldsymbol \beta}_\lambda$ is the least squares solution for any $\lambda <1$, the null vector for any $\lambda>1$, and any convex combination of the two if $\lambda=1$.
\end{theorem}
This theorem shows that using the $\ell_2$-norm for both the fit to the data and the penalty leads to a degenerate estimator (with an infinite number of solutions when $\lambda=1$, the fully sparse vector for all $\lambda>1$, and fitting the response exactly for $\lambda<1$). The theorem shows that SRAMlet is degenerate when $p=1$, which remains true in higher dimension.
See \citet{BLS14} for a study of the group square-root LASSO.

Consequently we use the LASSO penalty $P({\boldsymbol \beta})=\| {\boldsymbol \beta}\|_1$ for SRAMlet, in which case the estimator is not degenerate. 
First we consider square-root soft-waveshrink, that is, the univariate wavelet smoother defined as solution to \eqref{eq:sqrtAMlet} when $p=1$.
Square-root soft-waveshrink is the corner stone of SRAMlet.

\begin{definition} {\bf Square-root soft-waveshrink}. 
Given a response vector ${\bf y}^{(n)}$ corresponding to $n$ ordered univariate inputs, and an $n\times n$ orthonormal wavelet matrix $\Phi=[\Phi_0, \Psi]$, with $n_{\rm f}$ father wavelets $\Phi_0$ and $n_{\rm m}=(n-n_{\rm f})$ mother wavelets $\Psi$, and corresponding wavelet coefficients $({\boldsymbol \beta}_0, {\boldsymbol \beta})$, then, for a given positive penalty $\lambda$, the square-root soft-waveshrink wavelet coefficients estimates $(\hat {\boldsymbol \beta}_0, \hat {\boldsymbol \beta})$ are defined as a solution to
\begin{equation}\label{eq:SRSW}
\min_{{\boldsymbol \beta}_0, {\boldsymbol \beta}} \| {\bf y}^{(n)}-\Phi_0 {\boldsymbol \beta}_0-\Psi {\boldsymbol \beta}\|_2 + \lambda \| {\boldsymbol \beta}\|_1.
\end{equation}
\end{definition}

Square-root soft-waveshrink defined as a solution to~\eqref{eq:SRSW} has an implicit formulation via the soft-thresholding function, given in the following theorem.

\begin{theorem}
\label{th:closedform-1d}
The solution to~\eqref{eq:SRSW}
is  $\hat {\boldsymbol \beta}_0=\Phi_0^{\rm T}{\bf y}^{(n)}$ and 
$$
\hat {\boldsymbol \beta}=
\left\{
\begin{array}{rl}
{\bf z}; & \lambda \leq \frac{1}{\sqrt{\|{\bf z}\|_0}} \\
\eta_{\varphi(\hat {\boldsymbol \beta}; \lambda)}^{\rm soft}({\bf z}); & \frac{1}{\sqrt{\|{\bf z}\|_0}}< \lambda < \frac{\|{\bf z}\|_\infty}{\|{\bf z}\|_2}\\
{\bf 0}; & \lambda \geq \frac{\|{\bf z}\|_\infty}{\|{\bf z}\|_2}
\end{array}
\right. ,
$$
where ${\bf z}=\Psi^{\rm T}{\bf y}^{(n)}$, $\|{\bf z}\|_0=\lvert \{z_j \neq 0, j=1,\dots,n_{\rm m}\} \rvert$, the threshold function  is $\varphi(\cdot; \lambda)=\lambda \|{\bf z}-\cdot\|_2$, and the soft-thresholding function~\citep{Dono94b} is applied componentwise.
 To determine the implicit threshold $\varphi(\hat {\boldsymbol \beta}; \lambda)$, let
 $$
\varphi_{j}=\lambda \sqrt{\frac{\sum_{i=1}^{j}(|z|_{(i)})^2}{1-\lambda^2(n_{m}-j)}} \ \mbox{ for } \ j \in \{1,\ldots,n_{\rm m}-1\};
$$
then $\varphi(\hat {\boldsymbol \beta}; \lambda)=\varphi_{j^{*}}$ with $j^*=\left|\{i \in \{1,\ldots,n_{\rm m}\}: |z_i| \leq \varphi_{j^{*}} \}\right|=n_{\rm m}-\| \hat {\boldsymbol \beta}\|_0$.

\end{theorem}

Moreover to derive the Stein unbiased risk estimate for square-root soft-wavesrhink in Section~\ref{subsct:SURE}, we  need the following lemma.

\begin{lemma}
\label{LipschitzLemma}
 The function $\eta_{{\varphi({\boldsymbol \beta}; \lambda)}}^{\rm soft}({\bf z})$ of Theorem~\ref{th:closedform-1d} is Lipschitz continuous with respect to $({\bf z}, {\boldsymbol \beta})$.
\end{lemma}

A first consequence of Theorem~\ref{th:closedform-1d} is that
the estimate of standard deviation implicitly used by square-root soft-waveshrink  is
\begin{equation} \label{eq:implicitsigmahat}
 \hat \sigma=\varphi_{j^*}/\sqrt{n}/\lambda.
\end{equation}
So since  $\hat \sigma^2={\rm RSS}(\lambda)/n$,  square-root LASSO \citep{BCW11} and scaled LASSO \citep{scaledlasso12}, which idea was first introduced by \citet{articleAnestis10}, are equivalent. 

A second consequence of Theorem~\ref{th:closedform-1d} is that, owing to the fact that  ${\cal W}^{(n)}$ in \eqref{eq:orthoblock} is the concatenation of orthonormal blocks, the SRAMlet optimization problem~\eqref{eq:sqrtAMlet} can be solved by iteratively employing the solution to~\eqref{eq:SRSW}, as stated in the following theorem.

\begin{theorem}\label{th:convBCR}
The optimization \eqref{eq:sqrtAMlet} can be solved by block coordinate relaxation which consists in iteratively solving
$$
\min_{ {\boldsymbol \beta}_j\in {\mathbb R}^{n}}
\|{\bf y}^{(n)} - c {\bf 1}- {\cal W}^{(n)} {\boldsymbol \beta} \|_2 + 
\lambda \| {\boldsymbol \beta}\|_1, \quad {\rm where} \  {\boldsymbol \beta}=({\boldsymbol \beta}_1, \ldots, {\boldsymbol \beta}_p)
$$
for $j=1,\ldots,p$, then solving over $c\in {\mathbb R}$, and repeating until convergence.
\end{theorem}

\subsection{Selection of $\lambda$ by QUT}  \label{subsct:QUT}

In the spirit of the universal threshold of \citet{Dono94b} and  \citet{Dono95asym}, the first selection rule for $\lambda$ is the quantile universal threshold \citep{CaroNickJairoMe2017}. It is geared towards good identification of ${\cal S}$, and is based on the  property that the SRAMlet estimate \eqref{eq:sqrtAMlet} is the fully sparse zero-vector $\hat {\boldsymbol \beta}^{\rm SRAMlet}={\bf 0}$, given $\lambda$ is larger than a finite value that depends on the data. That specific value of $\lambda$ is given by the zero-thresholding function of Property~\ref{prop:ztf}.
\begin{property} \label{prop:ztf}
Given the matrix ${\cal X}^{(n)}$ and the output vector ${\bf y}^{(n)}$ of the training set, the smallest $\lambda$ for which $\hat {\boldsymbol \beta}^{\rm SRAMlet}$ solving \eqref{eq:sqrtAMlet} is the zero-vector is given by 
  the zero-thresholding function 
  \begin{equation}\label{eq:ztf}
   \lambda_0({\bf y}^{(n)}, {\cal X}^{(n)} )= \frac{\| {{\cal X}^{(n)}}^{\rm T}({\bf y}^{(n)}- \bar y^{(n)} {\bf 1}) \|_\infty}{\|({\bf y}^{(n)}- \bar y^{(n)} {\bf 1})\|_2}.
  \end{equation}
\end{property}

Under the assumption that all input entries carry no information, that is ${\cal S}=\emptyset$, the quantile universal threshold (QUT) selects $\lambda$ to be  large enough to satisfy $\hat {\boldsymbol \beta}^{\rm SRAMlet}={\bf 0}$ with probability $1-\alpha$, for a small $\alpha>0$, hence leading to $\hat {\cal S}=\emptyset$. So the selection of $\lambda$ is on a probabilistic scale governed by $\alpha$, in the spirit of hypothesis testing.
This selection rule, like the universal threshold of \citet{Dono95asym} is at the detection edge between signal and noise. For a given small $\alpha$, the QUT selection rule for $\lambda$ is defined below.

\begin{definition} \label{prop:QUT}
Given training inputs ${\cal X}^{(n)}$, let ${\bf Y}_0 \sim {\rm N}(c {\bf 1}, \sigma^2 I_n)$ be the distribution of ${\bf Y}$ according to \eqref{eq:YX} under the null model $H_0: \mu(x)=c$, and let $F_\Lambda$ be the c.d.f.~of $\Lambda = \lambda_0({\bf Y}_0, {\cal X}^{(n)}) $. For a small level $\alpha\in[0,1]$, the quantile universal threshold is defined as $\lambda_{{\rm QUT}}=F_\Lambda^{-1}(1-\alpha)$.
\end{definition}

Owing to the zero-thresholding function~\eqref{eq:ztf} that has both numerator and denominator proportional by multiplication of the response ${\bf y}^{(n)}$ by a scalar $\sigma$, the statistic $\Lambda$ is independent of~$\sigma$. Moreover substraction by $\bar y^{(n)}$ in both numerator and denominator makes the statistic $\Lambda$ independent of $c$. So the statistic $\Lambda$ is pivotal, and consequently, estimation of the noise standard deviation $\sigma$ is not required for SRAMlet for the selection of $\lambda$ by QUT. On the contrary, QUT for AMlet \citep{AMlet04} has the drawback of requiring an estimation of $\sigma$ because AMlet's zero-thresholding function is just the numerator of~\eqref{eq:ztf}.
And the estimation of $\sigma$ is a difficult problem in high dimension ($p$ large).
 An attempt for AMlet to circumvent this problem  consists in estimating $\sigma$ while iteratively solving the penalized least squares optimization problem. But this may lead to slow or even no convergence, because the AMlet optimization, although convex for a fixed $\lambda$, is no longer convex when the penalty (which depends on $\hat \sigma$) is regularly updated.
SRAMlet with QUT avoids the estimation of $\sigma$, which is a great advantage for additive models in high-dimension.

\subsection{Selection of $\lambda$ by SURE}  \label{subsct:SURE}

To select $\lambda$ with a good predictive performance measured by the mean squared error, also called $\ell_2$-risk, one can minimize over $\lambda$ an unbiased estimate of the risk \citep{Stein:1981}.

\begin{theorem}
\label{th:SURE}
The Stein unbiased risk estimate for square-root soft-waveshrink is
 $$
\mbox{{\rm SURE}}(\lambda)={\rm RSS}(\lambda)+2 \sigma^{2} n(\lambda) -n\sigma^{2},
$$
where ${\rm RSS}(\lambda)=\| {\bf y}^{(n)}-\Phi_0 \hat {\boldsymbol \beta}_0-\Psi \hat {\boldsymbol \beta}\|_2 ^{2}$ and $n(\lambda)= n_{\rm f}+\lvert \{\hat \beta^{\lambda}_j \neq 0, j=1,\dots,n_{\rm m}\} \rvert$.
\end{theorem}

As for soft-waveshrink, the degree of freedom of its square-root version is the number of non-zero coefficients. To use SURE, the estimation of $\sigma$ is needed; in dimension one, we recommend the MAD estimate of  \citet{Dono95i}.


\section{Monte Carlo simulation} \label{sct:simulation}
\subsection{Soft-waveshrink revisited}

Since square-root soft-waveshrink is the corner stone of SRAMlet, we first consider the univariate case $p=1$, and investigate the empirical properties of square-root soft-waveshrink in terms of mean squared error (MSE), true positive rate (TPR) and false discovery rate (FDR). We compare square-root soft-waveshrink (in black in the Figures) to the original soft-waveshrink (in red in the Figures).
It amounts to comparing square-root LASSO to LASSO in the orthonormal setting. 
To have an exactly  sparse representation of a function with wavelets, we consider the {\tt blocks} function with a signal to noise ratio (snr) equal to three \citep{Dono94b}  together with the use of Haar wavelets, all of which are piecewise constant functions on $[0,1]$. The number of father wavelets is $n_{\rm f}=2^3$ and the number of mother wavelets (that is, the potential needles) is $n_{\rm m}=n-8$. 

For each of $m=100$ Monte Carlo run, data ${\cal D}_k=\{(x_i, y_i)\}^{k}_{i=1,\ldots,n}$ for $k=1,\ldots,m$ are generated according to 
model~\eqref{eq:YX} with $\sigma=1$, $\mu=$ { \tt blocks} (with signal to noise ratio equal to three) sampled at $n=2^{10}$ random locations drawn from a uniform distribution on $[0,1]$. In dimension one, the needles ${\cal S}_k$ are defined as the non-zero mother wavelet coefficients obtained by applying the analysis wavelet operator to ${\boldsymbol \mu}^k=(\mu(x_{(1)}^k), \ldots, \mu(x_{(n)}^k))$, that is,  the true function $\mu$ evaluated at the ordered sampled locations $x_{(1)}^k, \ldots, x_{(n)}^k$, for $k=1,\ldots, m$.
Owing to the randomness of the sampled locations, the number of needles $s_k=|{\cal S}_k|$ varies; the  median value is of  $56$ needles from a total of $1016$ mother wavelets.

For the selection rule of $\lambda$, we consider three rules: oracle (that is, the $\lambda$ with the minimum $\ell_2$-loss),  SURE of Section~\eqref{subsct:SURE}, and QUT of Section~\eqref{subsct:QUT}. 
Two $\lambda_{\rm QUT}$ values are calculated for $\alpha=0.05$, one for soft-waveshrink and one for its square-root version. Soft-waveshrink requires an estimate for $\sigma$ that we take as the MAD estimate of \citet{Dono95i}. 

\begin{figure}[h]
\centering
\includegraphics[width=6.8in, height=5.7in]{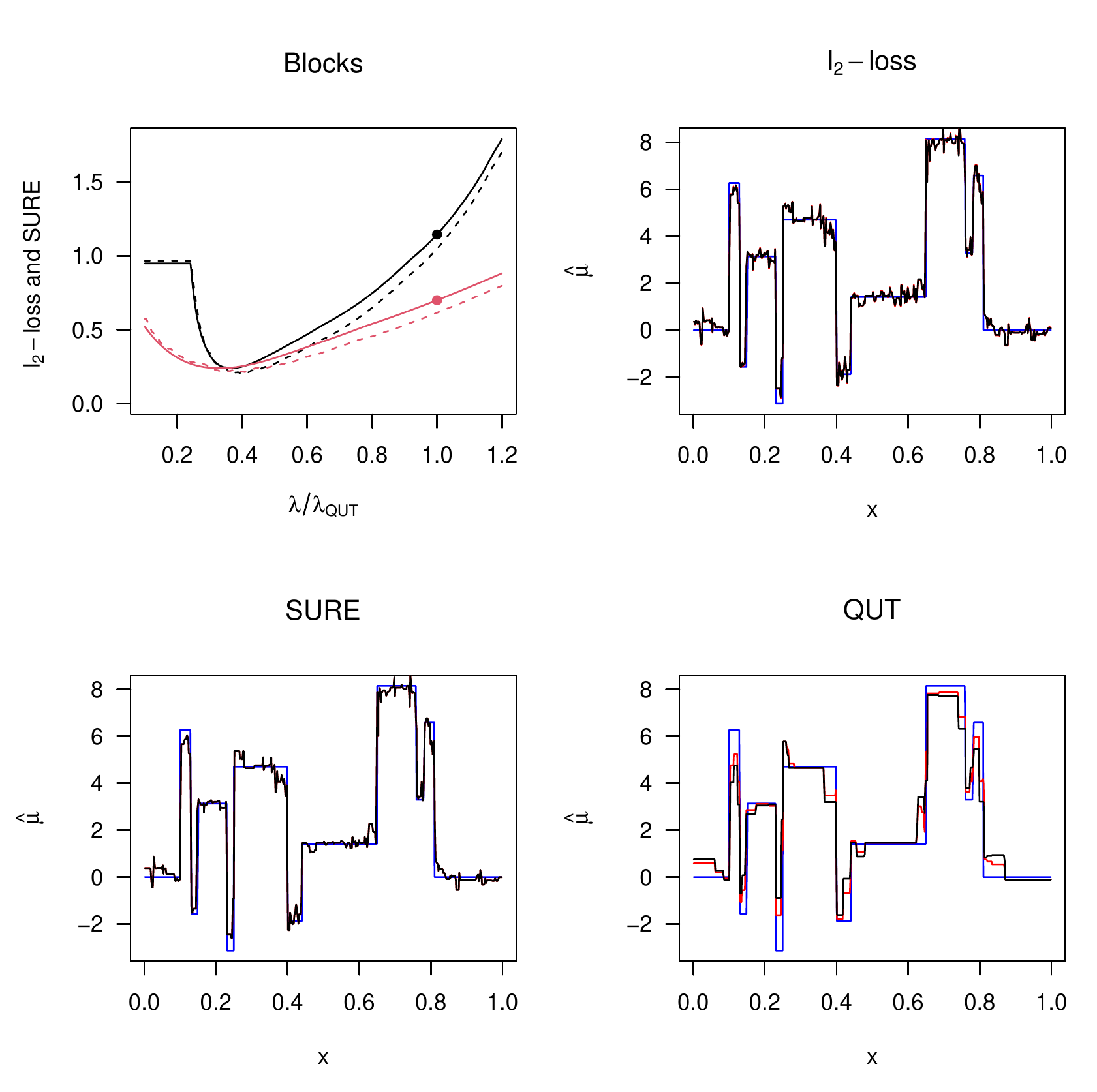}
\caption{Oracle $l_2$-loss (continuous) and SURE (dotted) for one sample of ``{\rm blocks}'' function sampled at n=$2^{10}$ equispaced points, for square-root (black) and original (red) soft-waveshrink.} 
\label{fig:2x2-loss}
\end{figure}

Figure~\ref{fig:2x2-loss} illustrates the differences between the three rules on a particular sample of the Monte Carlo simulation.
The top-left plot shows the $\ell_2$-loss (continous) and SURE (dotted) as a function of $\lambda/\lambda_{\rm QUT}$. 
The dots show the $\ell_2$-losses for $\lambda=\lambda_{\rm QUT}$. 
We observe that both SURE curves follow well their $\ell_2$-loss, and that $\lambda_{\rm QUT}$ is conservatively large, leading to a larger loss than the minimum of the curve.
The other three plots show the corresponding estimation of $\mu$, which corroborates that, with $\lambda_{\rm QUT}$, the estimation is less erratic than with SURE, which will translate to a lower rate of false jump detections.

%

\begin{figure}[h]
\centering
\includegraphics[width=6.8in, height=5.3in]{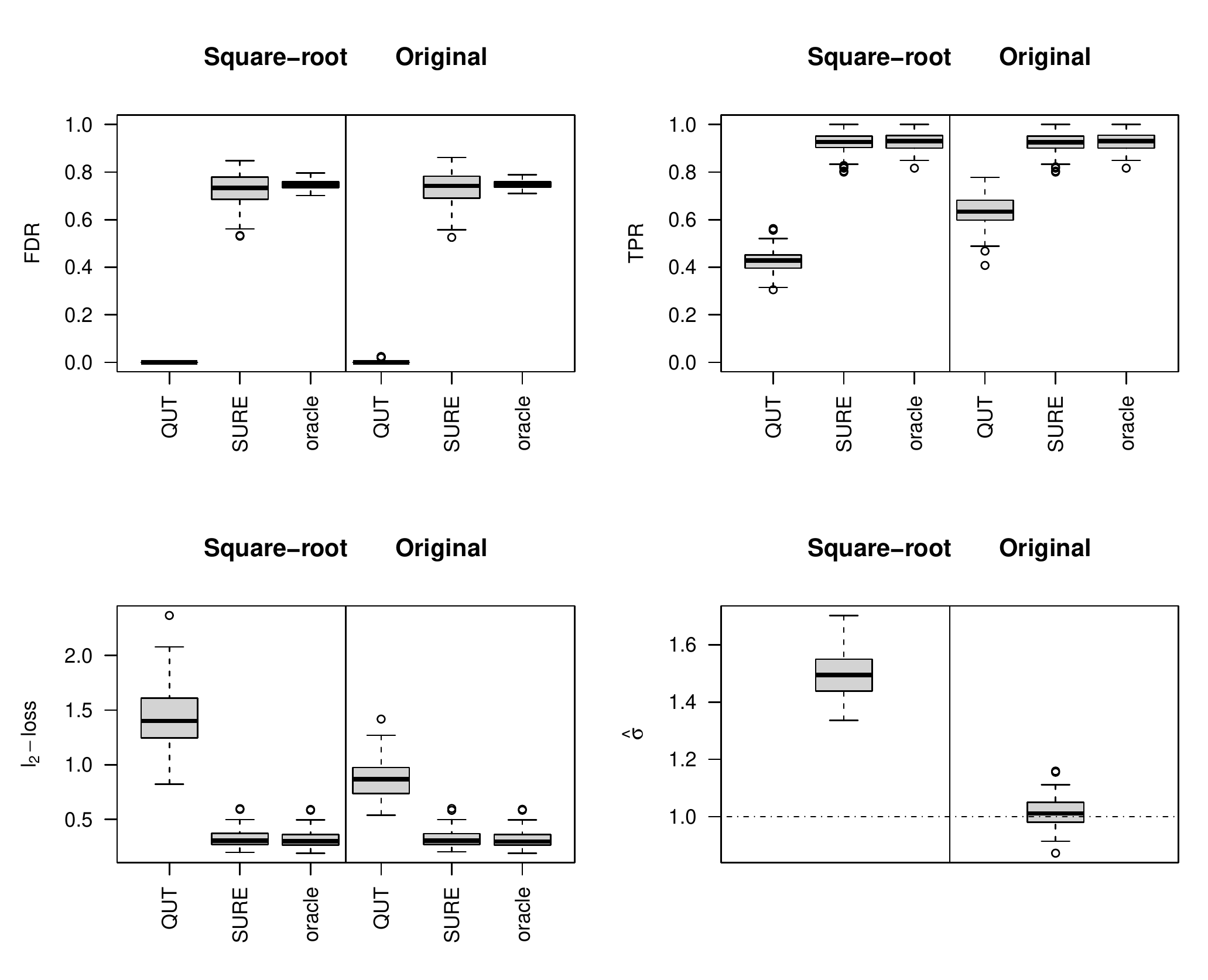}
\caption{The selection rule for $\lambda$  based on QUT, SURE and oracle.  Boxplots of $100$ Monte Carlo simulation runs for  square-root (half left) and original (half right) soft-waveshrink.
 FDR (top left), TPR (top right), $\ell_2$-loss (bottom left), and estimated $\sigma$'s (bottom right) with true $\sigma=1$ as 
the horizontal dotted line. Sample size $n=2^{10}$.  } 
\label{fig:2x2-tprfdr}
\end{figure}

Figure~\ref{fig:2x2-tprfdr}  summarizes the 100 Monte Carlo results with boxplots for FDR (top left), TPR (top right), $\ell_2$-loss (bottom left) and estimation of $\sigma$ (bottom right).  The best FDR--TPR trade--off \eqref{eq:FDRTPR} is with QUT, and the best $\ell_2$-loss is with oracle and SURE, which are both comparable. With QUT, the original soft-waveshrink has smaller  $\ell_2$-loss  than its square-root version thanks to a  smaller (and better) estimation of $\sigma$.
Indeed, as we can see on the bottom right plot,
for square-root soft-waveshrink, the implicit estimate given by \eqref{eq:implicitsigmahat} overestimates $\sigma$, while, for soft-waveshrink,  the MAD estimate of \citet{Dono95i} is centered around the true value $\sigma=1$.

%
%
%
%

\subsection{Sparse high dimensional additive models}

Because {\tt mgcv} builds spline matrices and selects many regularization parameters by minimizing GCV, it requires lots of memory and is computationally expensive. So we can only run small simulations with mgcv, that is $n$ and $p$ small.
And because \citet{HAM} search their hyperparameters on two grids that are not automatically calibrated to the data at hand, we are not able to run their method. Likewise  \citet{waveadditive2018} perform an expensive cross validation search and do not allow model selection.
We compare SRAMlet to the method of \citet{SAM} called Sparse Additive Models (SAM), and, to compare to a simple linear model, we also use the LASSO from the {\tt glmnet} library with the option '{\tt 1se}' for a conservative selection of $\lambda$ by cross validation.
When $n$ and $p$ are getting larger, only SRAMlet and AMlet can be employed since they do not require building and storing a large matrix in each direction. 

We choose to have $s=4$ out of $p$ inputs  with predictive information. 
Their corresponing functions $\mu_1, \ldots, \mu_4$ are {\tt blocks}, {\tt bumps}, {\tt heavisine} and {\tt Doppler} with ${\rm snr}=3$ \citep{Dono94b}.
The $p$ inputs are drawn from independent uniformally distributed random variable betwen zero and one. For wavelet-based methods, we use the  Daubechies ``extremal phase'' wavelets with a filter number equal to four and one father wavelet (the constant function that is not penalized).

For the first simulation, the number of samples is fixed to $n=2^{10}$ and the number of predictors $p\in\{10,100,1000\}$ varies. Table~\ref{tab:booktabs} reports the results in terms of MSE, FDR and TPR. First, we observe that all additive models (first four columns) perform better than the linear model (last column), as expected since the true association is additive, but not linear. Second, we see that only SRAMlet has the best low FDR--high TPR trade-off.

{\renewcommand\arraystretch{1.5}
\begin{table}[h]
\centering                           
  \resizebox{1\columnwidth}{!}{            
 \begin{tabular}{cc|c|c|c|c|c}
 \specialrule{0.1em}{0pt}{0pt}
 &&\multicolumn{4}{c|}{additive model} & \multicolumn{1}{c}{linear model} \\
 \cline{3-7}
 &{\bf p}&{\bf SRAMlet}&{\bf AMlet}$_{\hat \sigma}$&{\bf SAM}&{\bf MGCV}&${\bf LASSO}_{1se}$\\
\specialrule{0.1em}{0pt}{0pt}
\multirow{3}{*}{{\bf MSE}}&10&23.8(0.2)&22.0(0.2)&30.3(0.2)&27.5(0.2)&34.4(0.2)  \\ \cline{2-7}
&100&25.7(0.2)&22.6(0.2)&30.4(0.2)&/&34.3(0.2)  \\ \cline{2-7}
&1000&28.0(0.2)&24.0(0.2)&31.6(0.2)&/&35.2(0.2)  \\ \cline{2-7}
\specialrule{0.1em}{0pt}{0pt}
\multirow{3}{*}{\shortstack{{\bf FDR}}}&10 &0.07(0.01)&0.17(0.01)&0.48(0.02)&0.60(0)& 0.34(0.02) \\ \cline{2-7}
&100&0.12(0.02)&0.51(0.02)&0.79(0.01)&/&0.66(0.02)  \\ \cline{2-7}
&1000&0.14(0.02)&0.73(0.01)&0.85(0.01)&/& 0.70(0.03) \\ \cline{2-7}
\specialrule{0.1em}{0pt}{0pt}
\multirow{3}{*}{\shortstack{{\bf TPR}}}&10&1(0)&1(0)&0.99(0.003)&1(0)&0.82(0.02)  \\ \cline{2-7}
&100&0.99(0.005)&1(0)&0.98(0.007)&/&0.67(0.02)  \\ \cline{2-7}
&1000&0.94(0.01)&1(0)&0.89(0.02)&/&0.55(0.02)  \\ \cline{2-7}
\specialrule{0.1em}{0pt}{0pt}
   \end{tabular}}
   \caption{Empirical comparison of five methods based on the estimation of MSE, FDR and TPR by Monte-Carlo simulation. The size of the haystack is $p\in \{10,100,1000\}$, the number of needles is $s=4$ ({\tt blocks}, {\tt bumps}, {\tt heaviside}, {\tt Doppler}) and the sample size is fixed to $n=2^{10}$.}
   \label{tab:booktabs}
\end{table}}



\begin{figure}[h!]
\centering
\includegraphics[width=6.5in, height=5.6in]{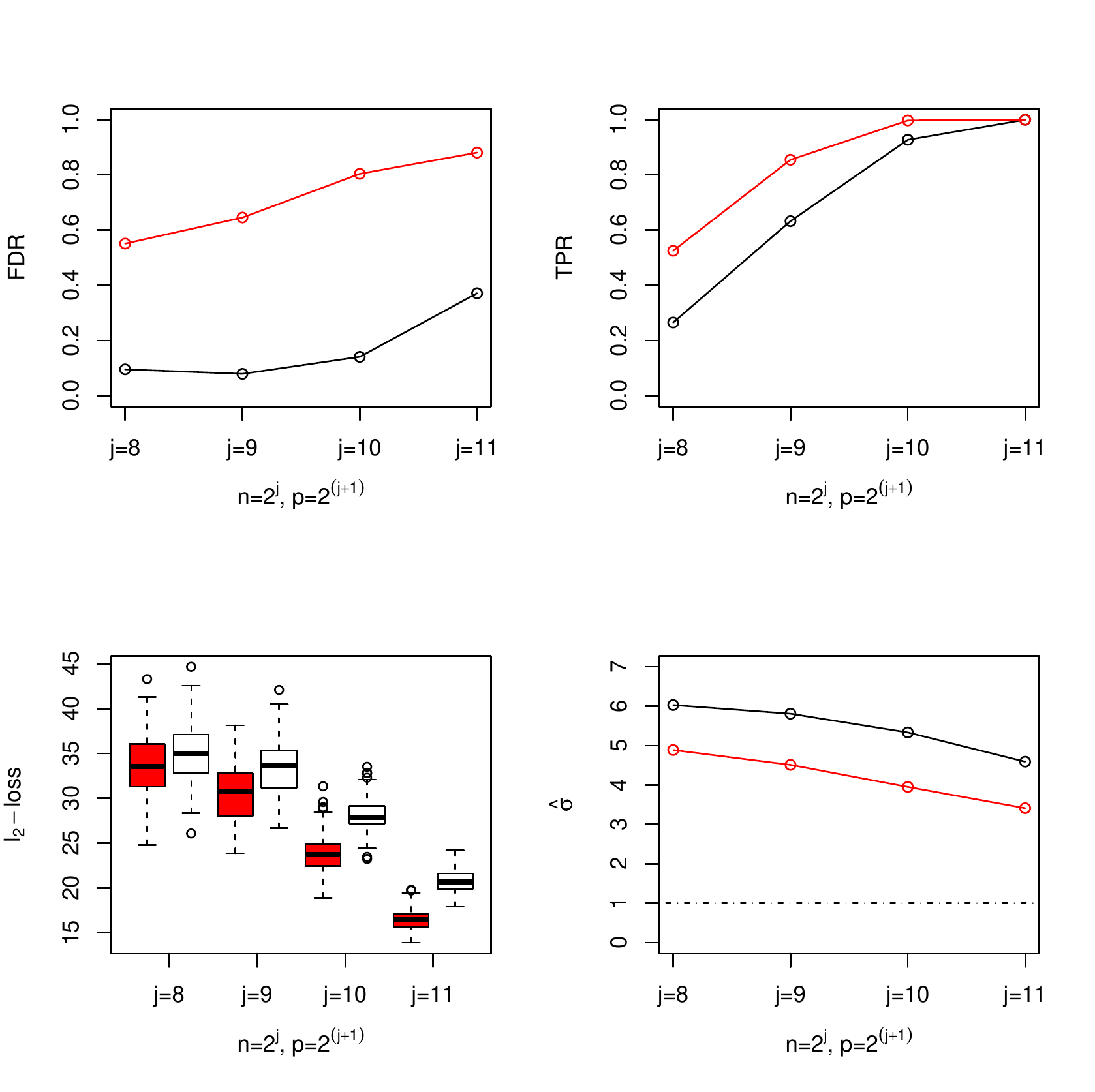}
\caption{Boxplots of $100$ Monte Carlo simulation runs for high-dimensional SRAMlet (black) and AMlet (red) with QUT for the selection of $\lambda$. Estimated $\sigma$ (top left) implicitely with SRAMlet and explicitely with AMlet,
$\ell_2$-loss (top right), FDR (bottom left) and TPR (bottom right).}
\label{fig:highdimension}
\end{figure}

For the second simulation, both $n$ and $p$ increase  with $(n,p)\in\{(2^j, 2^{j+1}),\ j\in \{8,9,10,11\}\}$. Because dimension are high (e.g., with $j=11$, the input matrix as $8'388'608$ entries), we can only apply the wavelet-based methods. Figure~\ref{fig:highdimension} summarizes the results. Again, we see that SRAMlet offers a good FDR--TPR trade--off.

Focusing on the estimation of $\sigma$, the bottom left plots of Figures~\ref{fig:2x2-tprfdr} and~\ref{fig:highdimension} show that the implicit estimation of $\sigma$ used by the square-root versions of soft-waveshrink and AMlet tends to be biased upwards, which can explain the good FDR control of the method. 

%
%

\section{Application to real data}

We consider three data sets of  sizes $n \in \{215,166,166\}$ and number of predictors  $p \in \{100,235,235\}$, respectively. We train the models on 128 observations for all three data sets, and test on the remaining data. We randomly sample from the original data twenty times and compute an average predictive error and average model size. We believe the underlying association is quite smooth so we use the Daubechies ``extremal phase'' wavelets with a filter number equal to four with periodic boundary correction. But as opposed to the Monte Carlo simulations, the underlying associations are not not necessarily periodic. To account for that, we use a model that concatenates three additive terms: a linear term, a smooth wavelet term, and a Haar wavelet term that does not suffer from boundary issues.
Tabel~\ref{tab:realdata} summarizes the results. We observe that SRAMlet selects the least inputs, but only with the {\tt ethanol} data (second column) does it achieve the best predictive error. Because $n$ is fairly small and $p$ is larger than $n$, the LASSO linear model is performing best in terms of predictive error, though with a fairly high number of selected inputs.

\section{Conlusion}

The wavelet-based SRAMlet method allows to fit sparse additive models in high dimensions, when both sample size $n$ and number of covariate $p$ are high. 
SRAMlet also offers a good complexity--predictive performance trade--off, as we observed on Monte-Carlo simulations based on the false discovery rate and true positive rate results.
SRAMlet can easily be robustified by Huberizing the least squares loss, which amounts to concatenating an identity matrix to the wavelet regression matrix \citep{Sardy01robustwavelet}.

{\renewcommand\arraystretch{1.5}
\begin{table}
\centering
\begin{tabular}{@{\extracolsep\fill}c|r|r|r|c@{\extracolsep\fill}}
\specialrule{0.1em}{0pt}{0pt}
\multicolumn{2}{c|}{\multirow{2}*{\diagbox[innerwidth=4cm]{Methods}{Datasets}}}&\multicolumn{1}{c|}{Meatspec}&\multicolumn{1}{c|}{Ethanol}&\multicolumn{1}{c}{Glucose} \\   \cline{3-5}
\multicolumn{2}{c|}{}&\multicolumn{1}{c|}{$n=215,p=100$}&\multicolumn{1}{c|}{$n=166,p=235$}&\multicolumn{1}{c}{$n=166,p=235$} \\   
\specialrule{0.1em}{0pt}{0pt}
\multirow{4}*{Model size}&SRAMlet&3(0.1)&5(0.2) &11(0.6)\\   \cline{2-5}
                                       &AMlet$_{\hat{\sigma}}$&12(2)&12(0.4)&16(0.4)\\   \cline{2-5}
                                       &SAM &53(3)&41(2)&58(7)\\\cline{2-5}
                                       &LASSO$_{1se}$&13(1)&10(0.5)&31(0.7)\\\cline{2-5}
                                       &$\overline{{\bf y}}_{training}$&0&0&0\\
\specialrule{0.1em}{0pt}{0pt}
\multirow{4}*{Predictive error}&SRAMlet&133(2)&2.7(0.2) &119(5)\\   \cline{2-5}
                                       &AMlet$_{\hat{\sigma}}$&143(3)&26(2) &68(5)\\   \cline{2-5}
                                       &SAM &37(4)&15(4)&46(7)\\\cline{2-5}
                                       &LASSO$_{1se}$&14(0.4)&3.2(0.1)&35(2)\\\cline{2-5}
                                       &$\overline{{\bf y}}_{training}$&169(4)&500(11)&196(7)\\
\specialrule{0.1em}{0pt}{0pt}
\end{tabular}
\caption{Results of accuracy versus mode size for five estimators (SRAMLET, AMlet, SAM, LASSO and null model) based on three real data sets.} \label{tab:realdata} %
\end{table}}

\appendix

\section{Proof of Theorem~\ref{th:sqrtgrpLASSO}} 
\label{app:sqrtgrpLASSO}

Consider the cost function $L(\boldsymbol{\alpha})=L_1(\boldsymbol{\alpha}) + L_2(\boldsymbol{\alpha})$ with  $L_1(\boldsymbol{\alpha})=\|{\bf y}-{\bf X}\boldsymbol{\alpha}\|_2$ and $L_2(\boldsymbol{\alpha})=\lambda\|\boldsymbol{\alpha}\|_2$, where $\lambda > 0, {\bf X} \text{ is orthonormal}$.
This fucntion $L$ is convex and, for  $\boldsymbol{\alpha}\neq {\bf 0} \text{ and } {\bf y}-{\bf X}\boldsymbol{\alpha} \neq {\bf 0}$, the first order optimality conditions are
\begin{equation*}
\begin{split}
\nabla_{\boldsymbol \alpha} L(\boldsymbol{\alpha})&=\frac{-{\bf X}^{T}({\bf y}-{\bf X}\boldsymbol{\alpha})}{\|{\bf y}-{\bf X}\boldsymbol{\alpha}\|_2}+\frac{\lambda\boldsymbol{\alpha}}{\|\boldsymbol{\alpha}\|_2}\equiv {\bf 0}
 \\ & \Leftrightarrow \boldsymbol{\alpha}= k(\boldsymbol{\alpha}) {\bf X}^{T}{\bf y} \text{ and } k(\boldsymbol{\alpha})=\frac{\|\boldsymbol{\alpha}\|_2}{\lambda \|{\bf y}-{\bf X}\boldsymbol{\alpha}\|_2+\|\boldsymbol{\alpha}\|_2}  
 \\ & \Leftrightarrow \boldsymbol{\alpha}= k(\boldsymbol{\alpha}) {\bf X}^{T}{\bf y} \text{ and } k(\boldsymbol{\alpha})= \frac{k(\boldsymbol{\alpha})\|{\bf y}\|_2}{\lambda \|{\bf y}-k(\boldsymbol{\alpha}) {\bf y}\|_2+k(\boldsymbol{\alpha}) \|{\bf y}\|_2} 
\\&  \Leftrightarrow  \boldsymbol{\alpha}= k({\boldsymbol \alpha}){\bf X}^{T}{\bf y}, (\lambda-1)(1-k({\boldsymbol \alpha}))=0, 0<k({\boldsymbol \alpha})<1\\
&  \Leftrightarrow  \boldsymbol{\alpha}= k{\bf X}^{T}{\bf y}, \lambda=1, 0<k<1.
\end{split}
\end{equation*}
So  a solution exists at $\boldsymbol{\alpha}=k{\bf X}^{T}{\bf y}$ for any $k\in(0,1)$ if and only if $\lambda=1$, in which case the cost is $L(k{\bf X}^{T}{\bf y})=\|{\bf y}\|_2$.
Moreover $\boldsymbol{\alpha}={\bf 0}$ is a solution if and only if
${\bf 0} \in \left\{ \nabla_{{\boldsymbol \alpha}}L_1({\bf 0})+\lambda\left\{{\bf g}:\|{\bf g}\|_2 \leq 1\right\}\right\}$,
where $\left\{{\bf g}:\|{\bf g}\|_2 \leq 1\right\}$ is the subgradient of $\|\boldsymbol{\alpha}\|_2$ at $\boldsymbol{\alpha}={\bf 0}$.
But $\| \nabla_{{\boldsymbol \alpha}}L_1({\bf 0}) \|_2=\| \frac{-{\bf X}^{T}{\bf y}}{\|{\bf y}\|_2} \|_2=1$ since $X$ is orthonormal. So  $\boldsymbol{\alpha}={\bf 0}$ is a solution if and only if $\lambda \geq 1$, in which case the cost is   $L({\bf 0})=\|{\bf y}\|_2$. 
Finally $\boldsymbol{\alpha}=X^{\rm T}{\bf y}$  is a solution if and only if $ {\bf 0} \in \left\{ \left\{{\bf g}:\|{\bf g}\|_2 \leq 1\right\} + \frac{\lambda\boldsymbol{\alpha}}{\|\boldsymbol{\alpha}\|_2}\right\}$,
where $\left\{{\bf g}:\|{\bf g}\|_2 \leq 1\right\}$ is the subgradient of $\|{\bf y}-{\bf X}\boldsymbol{\alpha}\|_2$ at $\boldsymbol{\alpha}=X^{\rm T}{\bf y}$. 
So  $\boldsymbol{\alpha}=X^{\rm T}{\bf y}$ is a solution if and only if  
 $\lambda \leq 1$, in which case the cost is $L(X^{\rm T}{\bf y})=\lambda\|{\bf y}\|_2$.
In summary, the solution is
$$
\boldsymbol{\alpha} = \left\{\begin{aligned}
{\bf X}^{T}{\bf y} & \qquad {\rm if }\ \lambda<1,\\
k{\bf X}^{T}{\bf y}, \ k\in(0,1) & \qquad {\rm if }\ \lambda=1,  \\
{\bf 0} & \qquad{\rm if }\  \lambda>1. 
\end{aligned}\right.
$$
%
%

\section{Proof of Theorem~\ref{th:closedform-1d}} 
\label{app:closedform-1d}

The optimization \eqref{eq:SRSW} is equivalent to 
\begin{equation}
\label{eq:SRSW1}
\min_{{\boldsymbol \beta}_0, {\boldsymbol \beta}} \left \| \left(\begin{array}{c} \Phi_{0}^{\rm T} {\bf y}^{(n)}\\ \Psi^{\rm T}  {\bf y}^{(n)} \end{array}\right)
-\left(\begin{array}{c} {\boldsymbol \beta}_0\\{\boldsymbol \beta}\end{array}\right) \right \|_2 + \lambda \left \| {\boldsymbol \beta} \right \|_1,
\end{equation}
thanks to the orthonormality of the wavelet matrix $\Phi$. 
Clearly the minimum with respect to ${\boldsymbol \beta}_0$ is at $\hat{{\boldsymbol \beta}}_0=\Phi_{0}^{T} {\bf y}^{(n)}$. Therefore we discuss the solution in ${\boldsymbol \beta}$ for ${\boldsymbol \beta}_0=\Phi_{0}^{T} {\bf y}$.

The cost function \eqref{eq:SRSW1} begin convex, we rely on
 Proposition~1.1 of \citet{DBLP:journals/corr/abs-1108-0775} to derive the solution in ${\boldsymbol \beta}$.  
First, letting ${\bf z}=\Psi^{\rm T}{\bf y}^{(n)}$, ${\boldsymbol \beta}={\bf 0}$ is the solution if the condition $\{{\bf 0}\} \in \left\{\frac{{\bf z}}{\|{\bf z}\|_2}+\lambda [-1,1]\cdot {\bf 1}\right\}$ is satisfied, which means  ${\bf -1}\leq\frac{{\bf z}}{\lambda\|{\bf z}\|_2} \leq {\bf 1}$. So only when $\lambda \geq \frac{\max_{i} |z_i|}{\|{\bf z}\|_2}=\frac{\|{\bf z}\|_\infty}{\|{\bf z}\|_2}$, can the condition  $\{{\bf 0}\} \in \left\{\frac{{\bf z}}{\|{\bf z}\|_2}+\lambda [-1,1]\cdot {\bf 1}\right\}$ hold. This is also consistent with the zero thresholding function \eqref{eq:ztf}.
Second, ${\bf z}$ is the solution if it satisfy $\{{\bf 0}\} \in \left\{ \left\{{\bf g} | \|{\bf g}\|_2 \leq 1\right\} + \lambda {\bf r} \right\}$, where ${\bf r}=[r_1,r_2,\ldots,r_{n_m}]^{T}$ and $r_i=\left\{ \begin{array}{rl}
 1, & z_i > 0 \\
 \left [-1,1\right ], & z_i=0 \\
-1,& z_i < 0 
\end{array}\right.$.
This condition means that 
$\lambda \leq \frac{1}{\|{\bf r}\|_2}$. With Gaussian  data, we have that $z_i \neq 0$ with probability one for $i=1,\ldots,n_m$, so $\|{\bf r}\|_2=\sqrt{n_m}$. But more generally, letting $\|{\bf z}\|_0$ be the number of nonzero in ${\bf z}$, then  $\|{\bf r}\|_2 \geq \|{\bf z}\|_0^{1/2}$. 
 So only when $\lambda \leq \frac{1}{\|{\bf z}\|_0^{1/2}}$, can the condition 
 $\{{\bf 0}\} \in \left\{ \left\{{\bf g} | \|{\bf g}\|_2 \leq 1\right\} + \lambda {\bf r} \right\}$ hold. In other words, $\lambda\|{\bf z}\|_0^{1/2}$ is the smallest shift of the ball $\left\{{\bf g} | \|{\bf g}\|_2 \leq 1\right\}$ that guarantees the condition to hold. If the shift is smaller or equal one, then the point ${\bf 0}$ still belongs to the ball.
Finally,  we consider when $\lambda \in (\frac{1}{\|{\bf z}\|_0^{1/2}},  \frac{\|{\bf z}\|_\infty}{\|{\bf z}\|_2})$. In that case, the KKT conditions are
$$
\{ 0\} \in \frac{\beta_i-z_i}{\|{\bf z}-\boldsymbol{\beta}\|_2}+\lambda 
\left\{ 
\begin{array}{rl}
 1, & \beta_i > 0 \\
 \left [-1,1\right ], & \beta_i=0 \\
-1,& \beta_i < 0 
\end{array}
\right. \quad i=1,\ldots,n_m,
$$
which is equivalent to
\begin{equation}\label{softsolution}
 {\boldsymbol \beta}=\eta^{\rm soft}_{\varphi(\boldsymbol{\beta}; \lambda)}(\bf z) \quad {\rm for} \quad \varphi(\boldsymbol{\beta}; \lambda)=\lambda \|{\bf z}-\boldsymbol{\beta}\|_2,
 \end{equation}
where $\eta^{\rm soft}_\varphi$ is the soft thresholding function with threshold $\varphi$ \citep{Dono94b}.
This is an implicit definition of the solution since ${\boldsymbol \beta}$ is in both the left and right hand sides of~\eqref{softsolution}.
 The number of zero entries in ${\boldsymbol{\beta}}$ is 
$$n_0({\boldsymbol{\beta}})=\left|\{i \in \{1,\ldots,n_m\}: \beta_i=0\}\right|=\left|\{i \in \{1,\ldots,n_m\}: |z_i| \leq \varphi({\boldsymbol{\beta}}; \lambda) \}\right|,$$ which satisfies $0<n_0({\boldsymbol{\beta}}) < n_m$ for the range of $\lambda$ considered. 
The threshold is also implicitely defined, but it is easy to see that 
$$
\varphi^2(\boldsymbol{\beta}; \lambda)=\lambda^2\left\{\varphi^2(\boldsymbol{\beta}; \lambda)(n_m-n_0(\boldsymbol{\beta}))+\sum_{i=1}^{n_0(\boldsymbol{\beta})}(|z|_{(i)})^2\right\},
$$
where $|z|_{(i)}$ is the $i$-th element for ordered $|{\bf z}|$. So among all 
$$
\varphi_{j}=\lambda \sqrt{\frac{\sum_{i=1}^{j}(|z|_{(i)})^2}{1-\lambda^2(n_m-j)}} \mbox{ for } j \in \{1,\ldots,n_m-1\},
$$
only $\varphi_{j^{*}}$ satisfying $j^*=n_0(\hat{\boldsymbol{\beta}})=\left|\{i \in \{1,\ldots,n_m\}: |z_i| \leq \varphi_{j^{*}} \}\right|=n_m-\|\hat{\boldsymbol{\beta}}\|_0$ leads to the solution $\hat{\boldsymbol{\beta}}=\eta_{\varphi_{j^*}}^{\rm soft}({\bf z})$. In practice, one tries all $j \in \{1,\ldots,n_m-1\}$ until $j=n_0(\eta_{\varphi_{j}}^{\rm soft}({\bf z}))$. 


\section{Proof of Lemma~1}
\label{Lipschitz}


\subsection{Lipschitz continuous with respect to \texorpdfstring{$\boldsymbol{\beta}$}{h}}

Given a fixed ${\bf z}\in {\mathbb R}^{n_{\rm m}}$ and $\lambda>0$, $\eta_{\lambda \|\boldsymbol{\beta}-{\bf z}\|_{2}}^{\rm soft}({\bf z})$ is Lipschitz continuous with respect to \texorpdfstring{$\boldsymbol{\beta}$}{h} if  there exists a constant $K_1$ such that
$$
\left\|\eta_{\lambda \|\boldsymbol{\beta}_1-{\bf z}\|_{2}}^{\rm soft}({\bf z})-\eta_{\lambda \|\boldsymbol{\beta}_2-{\bf z}\|_{2}}^{\rm soft}({\bf z})\right\|_2 
\leq K_1 \left\| \boldsymbol{\beta}_1-\boldsymbol{\beta}_2\right\|_2
$$
for all vectors $\boldsymbol{\beta}_1$ and $\boldsymbol{\beta}_2$.
Without loss of generality, we assume that $\lambda \|\boldsymbol{\beta}_1-{\bf z}\|_{2} \leq \lambda \|\boldsymbol{\beta}_2-{\bf z}\|_{2}$. For every element $z_i, i=1,\ldots, n_m$, we have the following three cases:
\begin{enumerate}
  \item if $z_i < -\lambda \|\boldsymbol{\beta}_2-{\bf z}\|_{2}$, then
  \begin{equation*}
  \begin{aligned}
  \left|\eta_{\lambda \|\boldsymbol{\beta}_1-{\bf z}\|_{2}}^{\rm soft}({z_i})-  \eta_{\lambda \|\boldsymbol{\beta}_2-{\bf z}\|_{2}}^{\rm soft}({z_i})\right|
  &=\left| z_i+ \lambda \|\boldsymbol{\beta}_1-{\bf z}\|_{2} -(z_i+\lambda \|\boldsymbol{\beta}_2-{\bf z}\|_{2})\right|\\
  & \leq \lambda\|\boldsymbol{\beta}_1-\boldsymbol{\beta}_2\|_{2} \mbox{ (Reverse triangle inequality)}.
  \end{aligned}
 \end{equation*}
Likewise when $z_i > \lambda \|\boldsymbol{\beta}_2-{\bf z}\|_{2}$ by  symmetry.
  
  \item if $-\lambda \|\boldsymbol{\beta}_2-{\bf z}\|_{2} \leq z_i < -\lambda \|\boldsymbol{\beta}_1-{\bf z}\|_{2}$, then
   \begin{equation*}
  \begin{aligned}
  \left|\eta_{\lambda \|\boldsymbol{\beta}_1-{\bf z}\|_{2}}^{\rm soft}({z_i})-  \eta_{\lambda \|\boldsymbol{\beta}_2-{\bf z}\|_{2}}^{\rm soft}({z_i})\right|
  &= \left| z_i-(- \lambda \|\boldsymbol{\beta}_1-{\bf z}\|_{2} )\right| \mbox{ (Distance)} \\
  & \leq  \left| - \lambda \|\boldsymbol{\beta}_2-{\bf z}\|_{2} -(- \lambda \|\boldsymbol{\beta}_1-{\bf z}\|_{2} )\right| \\
  &= \lambda \left|  \|\boldsymbol{\beta}_1-{\bf z}\|_{2}- \|\boldsymbol{\beta}_2-{\bf z}\|_{2} \right|
   \leq \lambda\|\boldsymbol{\beta}_1-\boldsymbol{\beta}_2\|_{2}.
  \end{aligned}
 \end{equation*}
 Likewise when $\lambda \|\boldsymbol{\beta}_1-{\bf z}\|_{2} < z_i \leq \lambda \|\boldsymbol{\beta}_2-{\bf z}\|_{2}$ by symmetry.

  \item if $-\lambda \|\boldsymbol{\beta}_1-{\bf z}\|_{2} \leq z_i \leq \lambda \|\boldsymbol{\beta}_1-{\bf z}\|_{2}$, then
   \begin{equation*}
  \begin{aligned}
  \left|\eta_{\lambda \|\boldsymbol{\beta}_1-{\bf z}\|_{2}}^{\rm soft}({z_i})-  \eta_{\lambda \|\boldsymbol{\beta}_2-{\bf z}\|_{2}}^{\rm soft}({z_i})\right|=
  \left|0-0\right|=0 \leq \lambda \|\boldsymbol{\beta}_1-\boldsymbol{\beta}_2\|_{2}.
  \end{aligned}
 \end{equation*}

\end{enumerate}
Putting the three cases together, we have that 
\begin{equation*}
\begin{aligned}
\left\|\eta_{\lambda \|\boldsymbol{\beta}_1-{\bf z}\|_{2}}^{\rm soft}({\bf z})-\eta_{\lambda \|\boldsymbol{\beta}_2-{\bf z}\|_{2}}^{\rm soft}({\bf z})\right\|_{2}^{2} &=\sum_{i=1}^{n_m}\left(\eta_{\lambda \|\boldsymbol{\beta}_1-{\bf z}\|_{2}}^{\rm soft}({z_i})-  \eta_{\lambda \|\boldsymbol{\beta}_2-{\bf z}\|_{2}}^{\rm soft}({z_i})\right)^{2}\\
&\leq \sum_{i=1}^{n_m} \lambda^2 \|\boldsymbol{\beta}_1-\boldsymbol{\beta}_2\|_{2}^{2}=n_m\lambda^2 \|\boldsymbol{\beta}_1-\boldsymbol{\beta}_2\|_{2}^{2}.
\end{aligned}
\end{equation*}  
So we get $
\left\|\eta_{\lambda \|\boldsymbol{\beta}_1-{\bf z}\|_{2}}^{\rm  soft}({\bf z})-\eta_{\lambda \|\boldsymbol{\beta}_2-{\bf z}\|_{2}}^{\rm soft}({\bf z})\right\|_{2} \leq K_1\|\boldsymbol{\beta}_1-\boldsymbol{\beta}_2\|_{2}$ for $K_1 = \lambda \sqrt{n_m}$.

 \subsection{Lipschitz continuous with respect to \texorpdfstring{${\bf z}$}{h}}
 
Given a fixed ${\boldsymbol{\beta}}\in {\mathbb R}^{n_{\rm m}}$  and $ \lambda>0$, $\eta_{\lambda \|\boldsymbol{\beta}-{\bf z}\|_{2}}^{\rm soft}({\bf z})$ is Lipschitz continuous with resect to ${\bf z}$ if there exists a constant $K_2$ such that
$$
\left\|\eta_{\lambda \|\boldsymbol{\beta}-{\bf z}_1\|_{2}}^{\rm soft}({\bf z}_1)-\eta_{\lambda \|\boldsymbol{\beta}-{\bf z}_2\|_{2}}^{\rm soft}({\bf z}_2)\right\|_2 
\leq K_2 \left\| {\bf z}_1-{\bf z}_2\right\|_2 
$$
for all vectors ${\bf z}_1$ and ${\bf z}_2$.
We consider two cases:
\begin{enumerate}
 \item if $\lambda \|\boldsymbol{\beta}-{\bf z}_1\|_{2} = \lambda \|\boldsymbol{\beta}-{\bf z}_2\|_{2}$. Consider now a particular $i\in \{1,\ldots,n_{\rm m}\}$ and, without loss of generality, suppose $z_{1i} < z_{2i}$. There care four sub-cases:
\begin{enumerate}
  \item if $z_{1i}<z_{2i}<-\lambda \|\boldsymbol{\beta}-{\bf z}_1\|_{2}$, then
  \begin{equation*}
  \begin{aligned}
  \left|\eta_{\lambda \|\boldsymbol{\beta}-{\bf z}_1\|_{2}}^{\rm soft}({ z}_{1i})-\eta_{\lambda \|\boldsymbol{\beta}-{\bf z}_2\|_{2}}^{\rm soft}({ z}_{2i})\right|
  &=\left|z_{1i}+ \lambda \|\boldsymbol{\beta}-{\bf z}_1\|_{2} -(z_{2i}+\lambda \|\boldsymbol{\beta}-{\bf z}_2\|_{2} )\right|\\
  &=\left|z_{1i}-z_{2i}\right| \leq \left|z_{1i}-z_{2i}\right|. 
  \end{aligned}
  \end{equation*}
  Likewise when $ \|\boldsymbol{\beta}-{\bf z}_1\|_{2} < z_{1i} < z_{2i}$ by symmetry.
    
  \item if $z_{1i}<-\lambda \|\boldsymbol{\beta}-{\bf z}_1\|_{2} \leq z_{2i} \leq \lambda \|\boldsymbol{\beta}-{\bf z}_1\|_{2} $, then
  \begin{equation*}
  \begin{aligned}
  \left|\eta_{\lambda \|\boldsymbol{\beta}-{\bf z}_1\|_{2}}^{\rm soft}({ z}_{1i})-\eta_{\lambda \|\boldsymbol{\beta}-{\bf z}_2\|_{2}}^{\rm soft}({ z}_{2i})\right|
  &=\left|z_{1i}-(-\lambda \|\boldsymbol{\beta}-{\bf z}_1\|_{2})\right|\\
  &\leq \left|z_{1i}-z_{2i}\right| \mbox{ (Distance)}. 
  \end{aligned}
  \end{equation*}
  
  Likewise when  $-\lambda \|\boldsymbol{\beta}-{\bf z}_1\|_{2} \leq z_{1i}\leq \lambda \|\boldsymbol{\beta}-{\bf z}_1\|_{2} < z_{2i}$ by symmetry.

  \item if $z_{1i}<-\lambda \|\boldsymbol{\beta}-{\bf z}_1\|_{2} < \lambda \|\boldsymbol{\beta}-{\bf z}_1\|_{2} <z_{2i}$, then
  \begin{equation*}
  \begin{aligned}
  \left|\eta_{\lambda \|\boldsymbol{\beta}-{\bf z}_1\|_{2}}^{\rm soft}({ z}_{1i})-\eta_{\lambda \|\boldsymbol{\beta}-{\bf z}_2\|_{2}}^{\rm soft}({ z}_{2i})\right|
  &=\left|z_{1i}-(- \lambda \|\boldsymbol{\beta}-{\bf z}_1\|_{2}) -(z_{2i}-\lambda \|\boldsymbol{\beta}-{\bf z}_2\|_{2})\right|\\
  &\leq \left|z_{1i}-(- \lambda \|\boldsymbol{\beta}-{\bf z}_1\|_{2})\right|+\left|z_{2i}-\lambda \|\boldsymbol{\beta}-{\bf z}_2\|_{2}\right|\mbox{ (Distance)}\\
  &\leq \left|z_{1i}-z_{2i}\right|  +  \left|z_{1i}-z_{2i}\right| =2\left|z_{1i}-z_{2i}\right|. 
  \end{aligned}
  \end{equation*}
 
 \item if $-\lambda \|\boldsymbol{\beta}-{\bf z}_1\|_{2} \leq z_{1i}< z_{2i} \leq \lambda \|\boldsymbol{\beta}-{\bf z}_1\|_{2}$, then
  \begin{equation*}
  \begin{aligned}
  \left|\eta_{\lambda \|\boldsymbol{\beta}-{\bf z}_1\|_{2}}^{\rm soft}({ z}_{1i})-\eta_{\lambda \|\boldsymbol{\beta}-{\bf z}_2\|_{2}}^{\rm soft}({z}_{2i})\right|
  &=\left|0-0\right|\leq\left|z_{1i}-z_{2i}\right|. 
  \end{aligned}
  \end{equation*}
  
  
\end{enumerate}
So $ \left|\eta_{\lambda \|\boldsymbol{\beta}-{\bf z}_1\|_{2}}^{\rm soft}({ z}_{1i})-\eta_{\lambda \|\boldsymbol{\beta}-{\bf z}_2\|_{2}}^{\rm soft}({z}_{2i})\right| \leq 2\left|z_{1i}-z_{2i}\right| $.

 \item if $\lambda \|\boldsymbol{\beta}-{\bf z}_1\|_{2}\neq \lambda \|\boldsymbol{\beta}-{\bf z}_2\|_{2}$. Following similar steps, we get $$ \left|\eta_{\lambda \|\boldsymbol{\beta}-{\bf z}_1\|_{2}}^{\rm soft}({ z}_{1i})-\eta_{\lambda \|\boldsymbol{\beta}-{\bf z}_2\|_{2}}^{\rm soft}({ z}_{2i})\right| \leq \max\left\{2\left|z_{1i}-z_{2i}\right|, \left|z_{1i}-z_{2i}\right|+\lambda \left\|{\bf z}_1-{\bf z}_2\right\|_{2}\right\}.$$
\end{enumerate}
Putting all results together, we get that
 \begin{equation*}
  \begin{aligned}
\left\|\eta_{\lambda \|\boldsymbol{\beta}-{\bf z}_1\|_{2}}^{\rm soft}({\bf z}_1)-\eta_{\lambda \|\boldsymbol{\beta}-{\bf z}_2\|_{2}}^{\rm soft}({\bf z}_2)\right\|_{2}^{2}
&=\sum_{i=1}^{n_m} \left(\eta_{\lambda \|\boldsymbol{\beta}-{\bf z}_1\|_{2}}^{\rm soft}(z_{1i})-\eta_{\lambda \|\boldsymbol{\beta}-{\bf z}_2\|_{2}}^{\rm soft}(z_{2i})\right)^{2}\\
& \leq \sum_{i=1}^{n_m} \left(\max\left\{2\left|z_{1i}-z_{2i}\right|, \left|z_{1i}-z_{2i}\right|+\lambda \left\|{\bf z}_1-{\bf z}_2\right\|_{2}\right\}\right)^2\\
&\leq \sum_{i=1}^{n_m} (2+\lambda )^2\left\|{\bf z}_1-{\bf z}_2\right\|_{2}^{2}
=(2+\lambda)^2 n_m \left\|{\bf z}_1-{\bf z}_2\right\|_{2}^{2}.
 \end{aligned}
 \end{equation*}

So $
\left\|\eta_{\lambda \|\boldsymbol{\beta}-{\bf z}_1\|_{2}}^{\rm soft}({\bf z}_1)-\eta_{\lambda \|\boldsymbol{\beta}-{\bf z}_2\|_{2}}^{\rm soft}({\bf z}_2)\right\|_{2} \leq K_2\|{\bf z}_1-{\bf z}_2\|_{2}$ with $K_2=(2+\lambda) \sqrt{n_m}$.

\subsection{Lipschitz continuous}

Given any vectors $({\bf z}_1^{\rm T},{\boldsymbol{\beta}}^{\rm T}_1)^{\rm T}$ and  $({\bf z}_2^{\rm T},{\boldsymbol{\beta}}_2^{\rm T})^{\rm T}$, we have 
\begin{equation*}
  \begin{aligned}
&\left\|\eta_{\lambda \|\boldsymbol{\beta}_1-{\bf z}_1\|_{2}}^{\rm soft}({\bf z}_1)-\eta_{\lambda \|\boldsymbol{\beta}_2-{\bf z}_2\|_{2}}^{\rm soft}({\bf z}_2)\right\|_2\\
&= \left\|\eta_{\lambda \|\boldsymbol{\beta}_1-{\bf z}_1\|_{2}}^{\rm soft}({\bf z}_1)-
\eta_{\lambda \|\boldsymbol{\beta}_2-{\bf z}_1\|_{2}}^{\rm soft}({\bf z}_1)+\eta_{\lambda \|\boldsymbol{\beta}_2-{\bf z}_1\|_{2}}^{\rm soft}({\bf z}_1)-
\eta_{\lambda \|\boldsymbol{\beta}_2-{\bf z}_2\|_{2}}^{\rm soft}({\bf z}_2)\right\|_2\\
&\leq  \left\|\eta_{\lambda \|\boldsymbol{\beta}_1-{\bf z}_1\|_{2}}^{\rm soft}({\bf z}_1)-
\eta_{\lambda \|\boldsymbol{\beta}_2-{\bf z}_1\|_{2}}^{\rm soft}({\bf z}_1)\right\|_2+ \left\|\eta_{\lambda \|\boldsymbol{\beta}_2-{\bf z}_1\|_{2}}^{\rm soft}({\bf z}_1)-
\eta_{\lambda \|\boldsymbol{\beta}_2-{\bf z}_2\|_{2}}^{\rm soft}({\bf z}_2)\right\|_2\\
&\leq K_1 \left\|\boldsymbol{\beta}_1-\boldsymbol{\beta}_2\right\|_2
+K_2  \left\|{\bf z}_1-{\bf z}_2\right\|_2\\
&\leq (K_1+K_2)\left\|\left(\begin{array}{cc} {\bf z}_1 \\{\boldsymbol{\beta}_1}\end{array}\right)-\left(\begin{array}{cc} {\bf z}_2 \\{\boldsymbol{\beta}_2}\end{array}\right)\right\|_2.
 \end{aligned}
 \end{equation*}
 So the function $\eta_{\lambda \|\boldsymbol{\beta}-{\bf z}\|_{2}}^{\rm soft}({\bf z})$ is Lipschitz continuous with respect to $({\bf z}, {\boldsymbol \beta})$.

\section{Proof of Theorem~\ref{th:convBCR}} 
\label{app:convBCR}

The SRAMlet optimization problem~\eqref{eq:sqrtAMlet} writes as
$$
\min_{c,{\boldsymbol \beta}} g(c,{\boldsymbol \beta})+\lambda \sum_{j=1}^p \sum_{i=1}^n h(\beta_{j,i}),
$$
where $g(c,{\boldsymbol \beta})=\|{\bf y}^{(n)}- c {\bf 1} - {\cal W}^{(n)} {\boldsymbol \beta} \|_2$ is a differentiable convex function on ${\mathbb R}^{1+pn}\backslash \{(c,{\boldsymbol \beta}):\ {\bf y}^{(n)}= c {\bf 1} + {\cal W}^{(n)} {\boldsymbol \beta} \}$ and $h(u)=|u|$ is a continuous function on ${\mathbb R}$. Under this separable structure of the non-differentiable part of the cost function, \citet{Tseng93} showed that the BCR algorithm using the systematic rule converges.

\section{Proof of Theorem~\ref{th:SURE}} 
\label{app:SURE}

Denote ${\bf r}=\Phi^{T}{\bf y}^{(n)}$, ${\boldsymbol{\theta}}=(\boldsymbol{\beta}_0^{T},\boldsymbol{\beta}^{T})^{T}$, ${\hat{\boldsymbol{\theta}}}=(\hat{\boldsymbol{\beta}}_0^{T},\hat{\boldsymbol{\beta}}^{T})^{T} $. First, we compute the gradient matrix $\frac{\partial \hat{\boldsymbol{\theta}}}{\partial {\bf r}}$. When $\lambda \leq \frac{1}{\sqrt{\|{\bf z}\|_0}}$, we get $\frac{\partial \hat{\boldsymbol{\theta}}}{\partial {\bf r}}={\bf I}_{n \times n}$.
When $\lambda \geq \frac{\|{\bf z}\|_\infty}{\|{\bf z}\|_2}$, we get $\frac{\partial \hat{\boldsymbol{\theta}}}{\partial {\bf r}}=\left(\begin{array}{cc}
{\bf I}_{n_f \times n_f} & {\bf 0}_{n_f \times n_m} \\
 {\bf 0}_{n_m \times n_f} & {\bf 0}_{n_m \times n_m} 
\end{array}\right)$.  For the third case $ \frac{1}{\sqrt{\|{\bf z}\|_0}} < \lambda < \frac{\|{\bf z}\|_\infty}{\|{\bf z}\|_2}$, 
the solution ${\boldsymbol \beta}$ in \eqref{softsolution} has an implicit function form, so we us the implicit function theorem to compute the gradient. To see this, consider the function
$f({\bf z},\boldsymbol{\beta})=\boldsymbol{\beta}-\eta_{\varphi(\boldsymbol{\beta}; \lambda)}^{\rm soft}({\bf z})$ that goes from $\mathbb{R}^{2n_m}$ to $\mathbb{R}^{n_m}$.
This  function is Lipschitz continuous from Lemma~\ref{LipschitzLemma}.  According to the Rademacher's theorem, the following Jacobian matrices exist almost everywhere.
For $i=1,\ldots,n_m$, we have
$$\left[\eta_{\varphi(\boldsymbol{\beta}; \lambda)}^{\rm soft}({\bf z})\right]_{i}=\left\{\begin{aligned} z_{i}+\lambda\|{\bf z}-\boldsymbol{\beta}\|_{2}; & \text {   if }-z_{i}>\lambda\|{\bf z}-\boldsymbol{\beta}\|_{2}\\ 0; & \text {   if }\left|z_{i}\right| \leq \lambda\|{\bf z}-\boldsymbol{\beta}\|_{2} \\ z_{i}-\lambda\|{\bf z}-\boldsymbol{\beta}\|_{2}; & \text {   if } z_{i}> \lambda\|{\bf z}-\boldsymbol{\beta}\|_{2} \end{aligned}\right. ,$$
\begin{equation*}
\begin{aligned}
\frac{\partial \left[\eta_{\varphi({\boldsymbol{\beta}}; \lambda)}^{\rm soft}({\bf z})\right]_{i}}{\partial z_i}=\left\{\begin{aligned} 1+\frac{\lambda(z_i-\beta_i)}{\|{\bf z}-{\boldsymbol{\beta}}\|_{2}}; & \text {   if }-z_{i}>\lambda\|{\bf z}-{\boldsymbol{\beta}}\|_{2}\\ 0; & \text {   if }\left|z_{i}\right| <\lambda\|{\bf z}-{\boldsymbol{\beta}}\|_{2} \\ 1-\frac{\lambda(z_i-\beta_i)}{\|{\bf z}-{\boldsymbol{\beta}}\|_{2}}; & \text {   if } z_{i}> \lambda\|{\bf z}-{\boldsymbol{\beta}}\|_{2} \end{aligned}\right.
=\left\{\begin{aligned} 1-\lambda^2; & \text {   if }|z_{i}|>\lambda\|{\bf z}-{\boldsymbol{\beta}}\|_{2}\\ 0; & \text {   if }\left|z_{i}\right| < \lambda\|{\bf z}-{\boldsymbol{\beta}}\|_{2}  \end{aligned}\right. ,
\end{aligned}
\end{equation*}
and for $j \neq i$,
\begin{equation*}
\begin{aligned}
\frac{\partial \left[\eta_{\varphi(\boldsymbol{\beta}; \lambda)}^{\rm soft}({\bf z})\right]_{i}}{\partial z_j}=\left\{\begin{aligned} \frac{\lambda(z_j-\beta_j)}{\|{\bf z}-\boldsymbol{\beta}\|_{2}}; & \text {   if }-z_j>\lambda\|{\bf z}-\boldsymbol{\beta}\|_{2}\\ 0; & \text {   if }\left|z_j\right| <  \lambda\|{\bf z}-\boldsymbol{\beta}\|_{2}\\ \frac{\lambda(\beta_j-z_j)}{\|{\bf z}-\boldsymbol{\beta}\|_{2}}; & \text {   if } z_j> \lambda\|{\bf z}-\boldsymbol{\beta}\|_{2} \end{aligned}\right. 
=\left\{\begin{aligned} -\lambda^2; & \text {   if }|z_{j}|>\lambda\|{\bf z}-\boldsymbol{\beta}\|_{2}\\ 0; & \text {   if }\left|z_{j}\right| <  \lambda\|{\bf z}-\boldsymbol{\beta}\|_{2}\end{aligned}\right. .
\end{aligned}
\end{equation*}
And for $i,j=1,\ldots, n_m$, we have
\begin{equation*}
\begin{aligned}
\frac{\partial \left[\eta_{\varphi({\boldsymbol{\beta}}; \lambda)}^{\rm soft}({\bf z})\right]_{i}}{\partial \beta_j}&=\left\{\begin{aligned} \frac{\lambda(\beta_j-z_j)}{\|{\bf z}-\boldsymbol{\beta}\|_{2}}; & \text {   if }-z_j>\lambda\|{\bf z}-\boldsymbol{\beta}\|_{2}\\ 0; & \text {   if }\left|z_j\right| < \lambda\|{\bf z}-\boldsymbol{\beta}\|_{2} \\ \frac{\lambda(z_j-\beta_j)}{\|{\bf z}-\boldsymbol{\beta}\|_{2}}; & \text {   if } z_j> \lambda\|{\bf z}-\boldsymbol{\beta}\|_{2} \end{aligned}\right. = \left\{\begin{aligned} \lambda^2; & \text {   if }|z_{j}|>\lambda\|{\bf z}-\boldsymbol{\beta}\|_{2}\\ 0; & \text {   if }\left|z_{j}\right| < \lambda\|{\bf z}-\boldsymbol{\beta}\|_{2}\end{aligned}\right. .
\end{aligned}
\end{equation*}

Denote 
$$
J_{\eta,{{\bf z}}}=\left(\begin{array}{cccc} \frac{\partial\left[\eta_{\varphi(\boldsymbol{\beta};\lambda)}^{\rm soft}({\bf z})\right]_1}{\partial z_1} & \frac{\partial\left[\eta_{\varphi(\boldsymbol{\beta};\lambda)}^{\rm soft}({\bf z})\right]_1}{\partial z_2} & \cdots& \frac{\partial\left[\eta_{\varphi(\boldsymbol{\beta};\lambda)}^{\rm soft}({\bf z})\right]_1}{\partial z_{n_m}} \\\frac{\partial\left[\eta_{\varphi(\boldsymbol{\beta};\lambda)}^{\rm soft}({\bf z})\right]_2}{\partial z_1} & \frac{\partial\left[\eta_{\varphi(\boldsymbol{\beta};\lambda)}^{\rm soft}({\bf z})\right]_2}{\partial z_2} & \cdots & \frac{\partial\left[\eta_{\varphi(\boldsymbol{\beta};\lambda)}^{\rm soft}({\bf z})\right]_2}{\partial z_{n_m}} \\ \vdots & \vdots& \ddots & \vdots \\\frac{\partial\left[\eta_{\varphi(\boldsymbol{\beta};\lambda)}^{\rm soft}({\bf z})\right]_{n_m}}{\partial z_1} & \frac{\partial\left[\eta_{\varphi(\boldsymbol{\beta};\lambda)}^{\rm soft}({\bf z})\right]_{n_m}}{\partial z_2} & \cdots & \frac{\partial\left[\eta_{\varphi(\boldsymbol{\beta};\lambda)}^{\rm soft}({\bf z})\right]_{n_m}}{\partial z_{n_m}}\end{array}\right),
$$

$$
J_{\eta,{{\boldsymbol{\beta}}}}=
\left(\begin{array}{cccc} \frac{\partial\left[\eta_{\varphi(\boldsymbol{\beta};\lambda)}^{\rm soft}({\bf z})\right]_1}{\partial \beta_1} & \frac{\partial\left[\eta_{\varphi(\boldsymbol{\beta};\lambda)}^{\rm soft}({\bf z})\right]_1}{\partial \beta_2} & \cdots& \frac{\partial\left[\eta_{\varphi(\boldsymbol{\beta};\lambda)}^{\rm soft}({\bf z})\right]_1}{\partial \beta_{n_m}} \\ \frac{\partial\left[\eta_{\varphi(\boldsymbol{\beta};\lambda)}^{\rm soft}({\bf z})\right]_2}{\partial \beta_1} & \frac{\partial\left[\eta_{\varphi(\boldsymbol{\beta};\lambda)}^{\rm soft}({\bf z})\right]_2}{\partial \beta_2} & \cdots & \frac{\partial\left[\eta_{\varphi(\boldsymbol{\beta};\lambda)}^{\rm soft}({\bf z})\right]_2}{\partial \beta_{n_m}} \\ \vdots & \vdots& \ddots & \vdots \\ \frac{\partial\left[\eta_{\varphi(\boldsymbol{\beta};\lambda)}^{\rm soft}({\bf z})\right]_{n_m}}{\partial \beta_1} & \frac{\partial\left[\eta_{\varphi(\boldsymbol{\beta};\lambda)}^{\rm soft}({\bf z})\right]_{n_m}}{\partial \beta_2} & \cdots & \frac{\partial\left[\eta_{\varphi(\boldsymbol{\beta};\lambda)}^{\rm soft}({\bf z})\right]_{n_m}}{\partial \beta_{n_m}}\end{array}\right).
$$

So the Jacobian matrix with respect to ${{\bf z}}$ is $J_{f,{\bf z}}=-J_{\eta,{{\bf z}}}$, the Jacobian matrix with respect to $\boldsymbol{\beta}$ is $J_{f,\boldsymbol{\beta}}={\bf I}_{n_m \times n_m}-J_{\eta,\boldsymbol{\beta}}$.
From the implicit function theorem, we  get 
\begin{equation*}
\begin{aligned}
\frac{\partial \hat{\boldsymbol{\beta}}}{\partial {{\bf z}}}&=-\left[J_{f,\boldsymbol{\beta}}\right]^{-1}\left[J_{f,{\bf z}}\right]\\
&=-\left[   {\bf I}_{n_m \times n_m}-J_{\eta,\boldsymbol{\beta}}  \right]^{-1} 
\left[ -J_{\eta,{{\bf z}}} \right]\\
&=\left[   {\bf I}_{n_m \times n_m}-J_{\eta,\boldsymbol{\beta}}  \right]^{-1} 
\left[ J_{\eta,{{\bf z}}} \right].
\end{aligned}
\end{equation*}
We notice that $\left[\frac{\partial {\hat{\boldsymbol{\beta}}}}{\partial {\bf z}}\right]_{[j,j]} = 1 \mbox{ if } |z_{j}|>\lambda\|{\bf z}-\boldsymbol{\beta}\|_{2}$, otherwise it is null. So the trace of $\frac{\partial {\hat{\boldsymbol{\beta}}}}{\partial {\bf z}}$ is $\lvert \{\hat \beta^{\lambda}_j \neq 0, j=1,\dots,n_{\rm m}\} \rvert$ when  $\frac{1}{\sqrt{\| {\bf z} \|_0}}<\lambda < \frac{\|{\bf z}\|_\infty}{\|{\bf z}\|_2}$.

We conclude the derivative of solution with data is
\begin{equation*}
\frac{\partial \hat{{\boldsymbol{\theta}}} }{\partial {\bf r}}=
\left\{
\begin{array}{cc}
{\bf I}_{n \times n}; & \lambda \leq \frac{1}{\sqrt{\|{\bf z}\|_0}}\\
\left(\begin{array}{cc}
{\bf I}_{n_f \times n_f } & {\bf 0}_{n_f  \times n_m} \\
 {\bf 0}_{n_m \times n_f} & {\bf 0}_{n_m \times n_m} 
\end{array}\right); & \lambda \geq \frac{\|{\bf z}\|_\infty}{\|{\bf z}\|_2}\\
\left(\begin{array}{cc}
{\bf I}_{n_f \times n_f} & {\bf 0}_{n_m \times n_m} \\
 {\bf 0}_{n_m \times n_f} &\left[   {\bf I}_{n_m \times n_m}-J_{\eta,{\boldsymbol{\beta}}}  \right]^{-1} 
\left[ J_{\eta,{\bf z}} \right]
\end{array}\right); & \frac{1}{\sqrt{\|{\bf z}\|_0}} < \lambda < \frac{\|{\bf z}\|_\infty}{\|{\bf z}\|_2}
\end{array}
\right. .
\end{equation*}

Stein's unbiased risk estimate formula leads to
\begin{equation*}
\begin{aligned}
\SURE(\lambda; (\hat {\boldsymbol \beta}_0, \hat {\boldsymbol \beta})^\lambda)&= \left\| \left(\begin{array}{c}
\hat{\boldsymbol{\beta}}_0^{\lambda} \\ \hat{\boldsymbol{\beta}}^{\lambda} \end{array}\right) - \left(\begin{array}{c}
\Phi_0^{T}{\bf y}^{(n)} \\ \Psi^{T}{\bf y}^{(n)}\end{array}\right) \right\|_2 + 2 \sigma^{2} \tr \left(  \frac{\partial \hat{{\boldsymbol{\theta}}} }{\partial {\bf r}}  \right)-n \sigma^{2}.
\end{aligned}
\end{equation*}
The first term is  ${\rm RSS}(\lambda)$ owing to the orthonormality of $\Phi$, and the second term is $n(\lambda)$.

\bibliographystyle{plainnat}
\bibliography{article_bis}

\begin{thebibliography}{29}
\providecommand{\natexlab}[1]{#1}
\providecommand{\url}[1]{\texttt{#1}}
\expandafter\ifx\csname urlstyle\endcsname\relax
  \providecommand{\doi}[1]{doi: #1}\else
  \providecommand{\doi}{doi: \begingroup \urlstyle{rm}\Url}\fi

\bibitem[Amato et~al.(2022)Amato, Antoniadis, De~Feis, and Gibels]{AnestisAM22}
U.~Amato, A.~Antoniadis, I.~De~Feis, and I.~Gibels.
\newblock Wavelet-based robust estimation and variable selection in
  nonparametric additive models.
\newblock \emph{Statistics and Computing}, 32:\penalty0 11, 2022.

\bibitem[Antoniadis(2010)]{articleAnestis10}
A.~Antoniadis.
\newblock Comments on: l1-penalization for mixture regression models.
\newblock \emph{TEST: An Official Journal of the Spanish Society of Statistics
  and Operations Research}, 19:\penalty0 257--258, 2010.

\bibitem[Bach et~al.(2011)Bach, Jenatton, Mairal, and
  Obozinski]{DBLP:journals/corr/abs-1108-0775}
F.~R. Bach, R.~Jenatton, J.~Mairal, and G.~Obozinski.
\newblock Optimization with sparsity-inducing penalties.
\newblock \emph{CoRR}, abs/1108.0775, 2011.

\bibitem[Barber and Cand\`es(2015)]{FDRCAndes2015}
R.~F. Barber and E.~J. Cand\`es.
\newblock Controlling the false discovery rate via knockoffs.
\newblock \emph{The Annals of Statistics}, 43\penalty0 (5):\penalty0
  2055--2085, 2015.

\bibitem[Belloni et~al.(2011)Belloni, Chernozhukov, and Wang]{BCW11}
A.~Belloni, V.~Chernozhukov, and L.~Wang.
\newblock Square-root lasso: pivotal recovery of sparse signals via conic
  programming.
\newblock \emph{Biometrika}, 98\penalty0 (4):\penalty0 791--806, 2011.

\bibitem[Buehlmann and van~de Geer(2011)]{BuhlGeer11}
P.~Buehlmann and S.~van~de Geer.
\newblock \emph{Statistics for {H}igh-{D}imensional {D}ata: {M}ethods, {T}heory
  and {A}pplications}.
\newblock Springer, Heidelberg, 2011.

\bibitem[Bunea et~al.(2014)Bunea, Lederer, and She]{BLS14}
F.~Bunea, J.~Lederer, and Y.~She.
\newblock The group square-root lasso: theoretical properties and fast
  algorithms.
\newblock \emph{IEEE Transactions on Information Theory}, 60\penalty0
  (2):\penalty0 1313--1325, 2014.

\bibitem[Daubechies(1992)]{Daub92}
I.~Daubechies.
\newblock \emph{Ten lectures on wavelets}.
\newblock Cambridge University Press, Philadelphia, 1992.

\bibitem[Donoho and Johnstone(1994)]{Dono94b}
D.~L. Donoho and I.~M. Johnstone.
\newblock Ideal spatial adaptation by wavelet shrinkage.
\newblock \emph{Biometrika}, 81\penalty0 (3):\penalty0 425--455, 1994.

\bibitem[Donoho and Johnstone(1995)]{Dono95i}
D.~L. Donoho and I.~M. Johnstone.
\newblock Adapting to unknown smoothness via wavelet shrinkage.
\newblock \emph{Journal of the American Statistical Association}, 90:\penalty0
  1200--1224, 1995.

\bibitem[Donoho et~al.(1995)Donoho, Johnstone, Kerkyacharian, and
  Picard]{Dono95asym}
D.~L. Donoho, I.~M. Johnstone, G.~Kerkyacharian, and D.~Picard.
\newblock Wavelet shrinkage: asymptopia?
\newblock \emph{Journal of the Royal Statistical Society: Series B},
  57\penalty0 (2):\penalty0 301--369, 1995.

\bibitem[Giacobino et~al.(2017)Giacobino, Sardy, Diaz~Rodriguez, and
  Hengartner]{CaroNickJairoMe2017}
C.~Giacobino, S.~Sardy, J.~Diaz~Rodriguez, and N.~Hengartner.
\newblock Quantile universal threshold for model selection.
\newblock \emph{Electronic Journal of Statistics}, 11\penalty0 (2):\penalty0
  4701--4722, 2017.

\bibitem[Haris et~al.(2018)Haris, Simon, and Shojaie]{waveadditive2018}
A.~Haris, N.~Simon, and A.~Shojaie.
\newblock Wavelet regression and additive models for irregularly spaced data.
\newblock In \emph{NeurIPS}, 2018.

\bibitem[Hastie and Tibshirani(1990)]{hastie1990generalized}
T.~Hastie and R.~Tibshirani.
\newblock \emph{Generalized additive models}.
\newblock Wiley Online Library, 1990.

\bibitem[Kerkyacharian and Picard(2004)]{KerkyacharianPicard04}
G.~Kerkyacharian and D.~Picard.
\newblock Regression in random design and warped wavelets.
\newblock \emph{Bernoulli}, 10:\penalty0 1053--1105, 2004.

\bibitem[Mallat(1989)]{MAllat89}
S.G. Mallat.
\newblock A theory for multiresolution signal decomposition: the wavelet
  representation.
\newblock \emph{IEEE Transactions on Pattern Analysis and Machine
  Intelligence}, 11\penalty0 (7):\penalty0 674--693, 1989.

\bibitem[Meier et~al.(2009)Meier, van~de Geer, and Buehlmann]{HAM}
L.~Meier, S.~van~de Geer, and P.~Buehlmann.
\newblock {High-dimensional additive modeling}.
\newblock \emph{The Annals of Statistics}, 37\penalty0 (6):\penalty0 3779 --
  3821, 2009.

\bibitem[Ravikumar et~al.(2009)Ravikumar, Lafferty, Liu, and Wasserman]{SAM}
P.~Ravikumar, J.~Lafferty, H.~Liu, and L.~Wasserman.
\newblock Sparse additive models.
\newblock \emph{Journal of the Royal Statistical Society: Series B},
  71\penalty0 (5):\penalty0 1009--1030, 2009.

\bibitem[Sardy and Tseng(2004)]{AMlet04}
S.~Sardy and P.~Tseng.
\newblock {AMlet}, {RAMlet} and {GAMlet}: automatic nonlinear fitting of
  {Additive Models}, {Robust} and {Generalized}, with wavelets.
\newblock \emph{Journal of Computational and Graphical Statistics},
  13:\penalty0 283--309, 2004.

\bibitem[Sardy et~al.(1999)Sardy, Percival, Bruce, and Gao]{SPBGS99}
S.~Sardy, D.~B. Percival, A.~G. Bruce, and W.~Gao, H-Y.and~Stuetzle.
\newblock Wavelet de-noising for unequally spaced data.
\newblock \emph{Statistics and Computing}, 9:\penalty0 65--75, 1999.

\bibitem[Sardy et~al.(2001)Sardy, Tseng, and Bruce]{Sardy01robustwavelet}
S.~Sardy, P.~Tseng, and A.~G. Bruce.
\newblock Robust wavelet denoising.
\newblock \emph{IEEE Transactions on Signal Processing}, 49:\penalty0
  1146--1152, 2001.

\bibitem[Stein(1981)]{Stein:1981}
C.~M. Stein.
\newblock Estimation of the mean of a multivariate normal distribution.
\newblock \emph{The Annals of Statistics}, 9\penalty0 (6):\penalty0 1135--1151,
  1981.

\bibitem[Sun and Zhang(2012)]{scaledlasso12}
T.~Sun and C.-H. Zhang.
\newblock Scaled sparse linear regression.
\newblock \emph{Biometrika}, 99:\penalty0 879--898, 2012.

\bibitem[Tibshirani(1996)]{Tibs:regr:1996}
R.~Tibshirani.
\newblock Regression shrinkage and selection via the lasso.
\newblock \emph{Journal of the Royal Statistical Society, Series B},
  58\penalty0 (1):\penalty0 267--288, 1996.

\bibitem[Tseng(1993)]{Tseng93}
P.~Tseng.
\newblock Dual coordinate ascent methods for non-strictly convex minimization.
\newblock \emph{Mathematical Programming}, 59:\penalty0 231--247, 1993.

\bibitem[Wahba(1990)]{Wahb:spli:1990}
G.~Wahba.
\newblock \emph{Spline Models for Observational Data}.
\newblock Society for Industrial and Applied Mathematics, Philadelphia, 1990.

\bibitem[Wood(2017)]{mgcv}
S.~N. Wood.
\newblock \emph{Generalized Additive Models: An Introduction with R}.
\newblock Chapman and Hall/CRC, 2 edition, 2017.

\bibitem[Wood et~al.(2016)Wood, Pya, and Saefken]{mgcvjasa}
S.~N. Wood, N.~Pya, and B.~Saefken.
\newblock Smoothing parameter and model selection for general smooth models.
\newblock \emph{Journal of the American Statistical Association}, 111\penalty0
  (516):\penalty0 1548--1563, 2016.

\bibitem[Yuan and Lin(2006)]{Yuan:Lin:mode:2006}
M.~Yuan and Y.~Lin.
\newblock Model selection and estimation in regression with grouped variables.
\newblock \emph{Journal of the Royal Statistical Society, Series B},
  68\penalty0 (1):\penalty0 49--67, 2006.

\end{thebibliography}

\end{document}